\documentclass[fleqn,usenatbib]{mnras}
\usepackage{newtxtext,newtxmath}
\usepackage[T1]{fontenc}
\DeclareRobustCommand{\VAN}[3]{#2}
\let\VANthebibliography\thebibliography
\def\thebibliography{\DeclareRobustCommand{\VAN}[3]{##3}\VANthebibliography}
\usepackage{graphicx}
\usepackage{amsmath}
\usepackage{longtable}
\usepackage{pdflscape}

\graphicspath{{images/}}

\title[Ex-situ Fractions of ETGs]{Mapping Accreted Stars in Early-Type Galaxies Across the Mass-Size Plane}

\author[T. Davison et al.]{
Thomas A. Davison,$^{1,2}$\thanks{E-mail: tdavison@eso.org}
Mark A. Norris$^{2}$,
Ryan Leaman$^{3,4}$,
Harald Kuntschner$^{1}$,
Alina Boecker$^{4}$,
\newauthor
Glenn van de Ven$^{3}$
\\
$^{1}$European Southern Observatory, Karl-Schwarzschild-Strasse 2, D-87548 Garching bei Muenchen, Germany\\
$^{2}$Jeremiah Horrocks Institute, University of Central Lancashire, Preston PR1 2HE, UK\\
$^{3}$Department of Astrophysics, University of Vienna, T\"urkenschanzstraße 17, A-1180 Vienna, Austria\\
$^{4}$Max-Planck-Institut f\"ur Astronomie, K\"onigstuhl 17, D-69117 Heidelberg, Germany
}

\date{Accepted XXX. Received YYY; in original form ZZZ}

\pubyear{2021}

\begin{document}
\label{firstpage}
\pagerange{\pageref{firstpage}--\pageref{lastpage}}
\maketitle

\begin{abstract}
Galaxy mergers are instrumental in dictating the final mass, structure, stellar populations, and kinematics of galaxies. Cosmological galaxy simulations indicate that the most massive galaxies at z=0 are dominated by high fractions of `ex-situ' stars, which formed first in distinct independent galaxies, and then subsequently merged into the host galaxy. Using spatially resolved MUSE spectroscopy we quantify and map the ex-situ stars in thirteen massive Early Type galaxies. We use full spectral fitting together with semi-analytic galaxy evolution models to isolate the signatures in the galaxies' light which are indicative of ex-situ populations. Using the large MUSE field of view we find that all galaxies display an increase in ex-situ fraction with radius, with massive and more extended galaxies showing a more rapid increase in radial ex-situ fraction, (reaching values between $\sim$30\% to 100\% at 2 effective radii) compared to less massive and more compact sources (reaching between $\sim$5\% to 40\% ex-situ fraction within the same radius). These results are in line with predictions from theory and simulations which suggest ex-situ fractions should increase significantly with radius at fixed mass for the most massive galaxies. 
\end{abstract}



\section{Introduction}\label{intro}
There are many indications from both observational and theoretical astrophysics that massive galaxies form `inside out'. This suggests that galaxies begin as a core of stars formed in-situ, and grow larger through both additional in-situ star formation and through the accretion of smaller galaxies \citep{Trujillo_2005, auger2011compact, perez2013evolution, van2015evidence}. As such, a gradient in stellar population parameters can be imprinted on a galaxy, with the largely in-situ core giving way to a more ex-situ dominated outskirts. Kinematic and population gradients have been found to exist frequently in galaxies \citep[e.g.][]{norris2006, naab2009minor, spolaor2010early, guerou2016exploring, Sarzi2018, Pinna2019, Dolfi2020, simonsgradient} however specific treatments to separate intrinsic gradients in the in-situ populations from those resulting from distributions of ex-situ material is a challenging task. This is largely a product of the difficulty in navigating around complex and interconnected secular processes that, additionally to accretion, can provide gradients in stellar populations and kinematics. As a result, disentangling the evidence of mergers is difficult to do via photometry, or from average metallicities and ages.

Evidence of this two-phase galaxy assembly also comes in the form of bimodality of globular star clusters (GCs) in colour and metallicity. GCs are found to display complex colour distributions in nearly every massive galaxy system studied \citep[e.g.][]{zinn1985globular, larsen2001properties, brodie2006extragalactic, yoon2006explaining, peng2006acs, villaume2019new, fahrion2020fornax}. Red globular clusters are thought to be closely linked to in-situ formation or very massive mergers with higher metallicities, whereas their blue counterparts are more indicative of acquisition from lower mass galaxies with lower metallicities. A result of this is that merging low-mass systems frequently provide GCs of low-metallicity, in stark contrast to in-situ metal rich GCs \citep{choksi2018formation, forbes2018}. These GC properties can be used to diagnose both merger history as well as gradients of ex-situ fraction \citep{forbes2015, kruijssen2018, beasley2018single, mackey2019}, however this is complicated by uncertain mappings between colour and metallicity, and uncertain ages of old GCs in systems outside the Local Group. 

Further evidence of a two-phase galaxy assembly scenario is derived from the faint stellar envelopes of massive galaxies. In \cite{huang2018individual} (and building on work from \citealt{huang2013fossil}) the authors use deep imaging to study the stellar halos of around 7000 massive galaxies from the Hyper Suprime-Cam (HSC) survey \citep{aihara2018first} out to $>$100kpc. The authors find that surface mass density profiles show relative homogeneity within the central 10-20kpc of the galaxies, however the scatter in this profile increases significantly with radius. Furthermore the authors find that the stellar halos become more prominent and more elliptical with increasing stellar mass. This is in line with a two-phase formation scenario in which central galaxy regions are formed by relatively stable in-situ processes, and the outskirts are formed through far more stochastic accretion and so show a greater scatter in the surface mass density. This is also found to be the case in \cite{oyarzun2019signatures} in which the authors find a flattening in the metallicity profile of z$<$0.15 early type galaxies beyond a radius of 1.5r$_e$, and conclude the most reasonable explanation of this is stellar accretion to the galaxy outskirts. This result is also seen for samples of brightest cluster galaxies \citep{edwards2020clocking} who likewise find signatures of the two-phase scenario in profiles of kinematics and metallicity.

Simulations of galaxy formation also show strong preferences for galaxies to evolve through frequent mergers, and by accreting material to their outskirts \citep{kobayashi2004grape, zolotov2009, oser2010two, navarro2013, van20143d, rg2016}. This is particularly clear for the most massive ellipticals which show strong gradients of increasing ex-situ fraction with galactocentric radius, as well as high total fractions of ex-situ stars (with galaxies of mass M$_{\odot}$ $>$ 1.7$\times$10$^{12}$ M$_{\odot}$ being composed of populations with an ex-situ fraction of $\approx$90\%) \citep{oser2010two,lackner2012building,rg2016,pillepich2017first,davison2020eagle}. Despite clear trends in ex-situ fraction, these galaxies show a strong overlap between in- and ex-situ populations within 2 effective radii \citep{pillepich2015building} which causes difficulty in photometric approaches to ex-situ population extraction \citep[see also][]{remus2021accreted}. Modern photometric methods for identification of accreted stars and signatures of interaction (such as tidal features) are able to accurately identify stellar features in the stellar halo \citep[e.g.][]{duc2015atlas3d, morales2018systematic, martinez2021hidden}. Advanced deep-imaging methods can identify features of interaction out to 10 effective radii \citep{jackson2021nature}. Despite these ongoing advances in treatments of photometric data, most have difficulty accurately quantifying ex-situ fractions in the centres of galaxies ($<$2r$_e$) especially for ancient mergers which have largely diffused in projection space.  

Notably, galaxy size seems to be closely linked to galaxy accretion history. Literature has widely shown that galaxies at high redshift are on average far more compact than similar galaxies in the local universe \citep{van2006space, bezanson2009relation, belli2015stellar, wellons2016diverse}. Larger galaxies are expected to grow radially as a result of dry-mergers disproportionately extending massive galaxies \citep{cappellari2013effect, barro2013candels}. For this reason it can be highly instructive to examine galaxy properties related to accretion in both terms of galaxy radius and galaxy mass.

Analyses of the Illustris simulation \citep{genel2014introducing, vogelsberger2014introducing, vogelsberger2014properties} have shown the impact of mergers on galaxy populations, and work by \cite{cook2016information} further explored the population gradients in Illustris galaxies. These show clear demonstrations that large galaxies gain the majority of their material from ex-situ sources, rather than in-situ star formation. Particularly, \cite{rg2016} find a cross-over point for nearby galaxies (z$<$1, M$_*\approx$1-2$\times$10$^{11}$M$_{\odot}$) wherein galaxies transition from in-situ dominated growth to ex-situ dominated growth. For all galaxies, the authors find a roughly even split of merger mass contribution between major mergers (mass ratio $>$1/4) and minor/very minor mergers (mass ratio $<$1/4). 

This is similarly shown for the Magneticum simulations in \cite{remus2021accreted} where the mean Magneticum galaxy ex-situ fraction passes 50\% at M$_*\approx$8$\times$10$^{10}$M$_{\odot}$. Here the authors find similar splits in galaxy mass contribution from major and minor mergers, though with different defined mass splits. They show that the major merger mass contribution is consistently larger than minor mass contribution, but this is dependent on galaxy `class' (with division based upon radial stellar density profiles), see their figure 6 for detail.

Similarly in analyses of the EAGLE simulations \citep{schaye2015eagle, crain2015eagle}, almost all galaxies are found to contain a non-negligable quantity of accreted stellar mass \citep{davison2020eagle}. Clearly the influence of ex-situ populations from mergers and fly-bys on galaxy evolution is profound, influencing the composition, kinematics and star formation mechanisms of a galaxy. Stellar material acquired in mergers has been shown to be the key contributor to stellar mass in massive galaxies. As shown in \cite{davison2020eagle}, the lowest mass galaxies analysed (M$_*\approx$1$\times$10$^{9}$M$_{\odot}$) comprise of 10$\pm$5\% ex-situ stars. This increases with mass up to the most massive galaxies in the simulation (M$_*\approx$2$\times$10$^{12}$M$_{\odot}$) which are comprised of 80$\pm$9\% ex-situ stars. Interestingly this work highlighted trends in ex-situ fraction with galaxy density, showing that at fixed mass more extended galaxies would on average contain higher fractions of ex-situ stars.  

With the recent advancement in integral field spectroscopy, galaxies are being studied spectroscopically as spatially resolved objects, detailing spectral differences with galactocentric radius, and physical location within a galaxy \cite[see e.g.][]{guerou2016exploring, mentz2016abundance}. Instruments such as SAURON at the WHT, GCMS (VIRUS-P) on the 2.7m Harlan J. Smith telescope and MUSE (Multi-unit spectroscopic explorer) at the VLT \citep{sauron1,hill08,bacon2010muse} with their $\sim$1 arcminute field of views have been critical to the development of this field, and are a few of the integral field units (IFUs) driving this particular area of research. IFUs have significantly widened the field of galactic archaeology for nearby galaxies as they have allowed for more thorough spatial investigations of population distributions. Derived population maps can provide powerful insights into visual features, kinematics, and evolution \cite[see e.g.][]{comeron2015, faifer2017, Ge2019, davison2021old}. 

Alongside these advancements in instrumentation are equally important advancements in software. Full spectral fitting has evolved to fit the needs of the extraordinary data taken by IFUs. A number of full spectral fitting codes now exist which can efficiently reconstruct stellar populations present within galaxy spectra. Examples of such software are pPXF \citep{cappellari2004parametric, cappellari2016improving}, FIREFLY \citep{wilkinson2017firefly}, STECKMAP \citep{ocvirk2006steckmap}, and STARLIGHT \citep{cid2005stellar}. Full spectral fitting software combines single-age single-metallicity stellar spectral models (single stellar population models) in a weighted grid in order to reproduce a provided spectrum. A linear least squares algorithm can determine the optimal stable solution of model combinations to reproduce an input spectrum. Template regularisation provides dampening of noise, and allows for a smoothed physical solution which better represents the distribution of ages and metallicity in a single integrated spectrum. With application to resolved spectral datasets of galaxies received from IFUs, it has become possible to extract maps of stellar populations for a given galaxy. This has provided remarkable insight into the stellar kinematic and population properties of galaxies \cite[see e.g.][for examples of kinematic and population analysis from full spectral fitting]{Onodera12,norris2015extended,ferre-mateu17,Kacharov18,ruiz-lara18, boecker2020recovering}.

A method to identify ex-situ stellar populations by exploiting these spectroscopically recovered age-metallicity distributions was proposed in \cite{boecker2020galaxy}. In this method the age-metallicity distribution of each galaxy is determined via full spectral fitting. Chemical evolution templates for galaxies of different masses are calculated, depicting the in-situ star formation and chemical enrichment history of a galaxy. Lower mass galaxies typically have lower metallicities at fixed age than those at higher mass, and as such ex-situ stars can be separated from the recovered age-metallicity distribution of the host galaxy. This technique was applied to mock spectra from simulated galaxies from the EAGLE cosmological simulation suite and demonstrated remarkably accurate results, such that the spectroscopically recovered accretion fractions and merger histories matched the known merger history of individual simulated galaxies.

Literature examining the stellar population profiles of galaxies has proven to be particularly useful for exploring galaxy evolution when considering the two-phase formation scenario. The MASSIVE survey \citep{ma2014massive} has provided insights into the links between stellar populations of galaxies, and their kinematics and abundance ratio patterns, finding a link between radial anisotropy and metal-poor populations, most likely as a result of minor mergers and accretion \citep{greene2013stellar, greene2015massive, greene2019massive}. Further population based evidence comes from the MaNGA survey \citep{zheng2017sdss, goddard2017sdss, li2018sdss}. In \cite{oyarzun2019signatures} the authors find variable metallicity gradients and flattening in the outskirts of MaNGA galaxies, suggesting that this is a sign of stellar accretion, and these features are also found in simulated environments \cite[see e.g.][]{taylor2017metallicity}. The SAMI survey has likewise provided clues of accretion from stellar population profiles \citep{scott2017sami, ferreras2019sami, santucci2020sami}.

Work in \cite{spavone2021assembly} has utilised surface brightness distributions of MUSE galaxies to examine assembly history, with 3 massive galaxies ($M_* > 10^{12}M_{\odot}$) shown to have ex-situ fractions of $>$77\%. They further suggest that the majority of this ex-situ mass is obtained via major mergers. The work utilises deep imaging in combination with MUSE data in order to speculate on the merger mechanisms and accreted fractions of galaxies, however MUSE coverage is limited to within $\sim 1R_e$.

In this paper we leverage high quality MUSE IFU data along with new analytic models and full spectral fitting techniques to measure the radial variation in ex-situ fraction across the galaxy mass-size plane. We improve and expand upon the \cite{boecker2020galaxy} method and chemical evolution templates to allow for a more sophisticated ex-situ fraction determination and uncertainty quantification. Ascertaining how ex-situ fraction correlates with galaxy density will provide indirect understanding of the impact of dry mergers on the host structure as well as stellar populations. Analytic expectations for the size-growth due to dry minor mergers would naively suggest that the least dense massive ellipticals may show stronger stellar population signatures of ex-situ accretion.

We present the analysis of 13 resolved galaxies observed by MUSE. In order to estimate ex-situ fractions of these galaxies, full spectral fitting (Section \ref{fsf}) is used to quantify the ages and metallicities of stars present in each binned area of the galaxy data-cube. Following this, the in-situ and ex-situ stellar components are separated using analytic prescriptions for the expected galaxy assembly pathways so that we may estimate the quantity of ex-situ stars present in a galaxy (Section \ref{exfrac}).


The layout of the paper is as follows. In Section \ref{method} we explain in detail the methodology used to gather and process data, including binning methods and merger modelling. We present the results of this methodology in Section \ref{results}, and discuss the implication of the results in Section \ref{discussion}. Finally in Section \ref{conclusion} we provide concluding remarks.

\section{Methodology}\label{method}
A galaxy sample was obtained by first searching for all galaxies that have been observed with MUSE. A visual inspection of the dataset showed that lower mass resolved galaxies were poorly represented by existing MUSE data. With this in mind a primary mass selection was defined, limiting galaxies to between 1$\times10^{11}<M_{\odot}<2.5\times10^{12}$ in order to facilitate examining galaxy features with respect to galaxy size at fixed mass. This was then cut to those galaxies with accurate distance estimates (via SN1a, TRGB, or GCLF), and further cut to galaxies with good imaging data, including estimates of effective radius and ellipticity. Requirements were also imposed to only consider targets with MUSE spatial coverage to a minimum of 1.5 effective radii (though preferably to 3), high spatial resolution, and quality data cubes without reduction errors. The distribution of these galaxies against a wider sample is shown in Figure \ref{subsel}. These selected MUSE data cubes were obtained from the ESO MUSE archive from multiple existing sources. All final cube mosaics include the galaxy core and extend contiguously to the outer regions. Data-cubes were obtained from the following ESO programs: 094.B-0298 (P.I. Walcher, C.J), 296.B-5054 (P.I. Sarzi, M), 0103.A-0447 \& 097.A-0366 (P.I. Hamer, S), 60.A-9303 (SV), 094.B-0321 P.I. (Marconi, A), 094.B-0711 (P.I. Arnaboldi, Magda). Data were taken as pre-reduced data-products in the majority of cases. In some cases where cubes were reduced poorly or erroneously, cubes were re-reduced using the standard ESO pipeline \citep{weilbacher2020data}. Reduction applied standard MUSE calibration files and all exposures were reduced with bias subtraction, flat-fielding, and were wavelength calibrated.

Galaxy mass was estimated using relations between J-K colour and K-band stellar mass-to-light-ratios ($\Upsilon^{'}_K $). J-K and K-band magnitudes were provided by the 2Mass Large Galaxy Atlas \citep{skrutskie2006two}. As shown in \cite{westmeier2011gas}, using analysis from \cite{bell2001stellar}, the J-K to M/L relation is defined as:
\begin{equation} 
log_{10}\left(\Upsilon^{'}_K\right)=1.434\left(J-K\right)-1.380
\end{equation}
and assumes a modified Salpeter IMF \citep{salpeter1955luminosity, fukugita1998cosmic}.

These stellar mass-to-light-ratios were combined with K-band magnitudes and errors therein to produce mass estimates. Size was derived from 2MASS LGA J-band half-light radii estimates, in combination with distance estimates. Monte-Carlo simulation of magnitude, distance, apparent radius, and other uncertainties was employed to estimate final uncertainty in mass and size. For each galaxy, Monte-Carlo simulations were run 15000 times, considering all sources of error to produce a probability distribution of mass and size. Distance estimates and errors impacted on the calculation of physical radius, which in turn impacted on the area with which to calculate mass using mass to light ratios.

A full process example Figure can be found in Figure \ref{1407_poster} which shows the reduction and analysis for NGC\,1407. 

\subsection{Observations and Analysis}
For all binning methods, a `first sweep' examined the signal-to-noise (S/N) of all spaxels in the datacube. The S/N of each spaxel was defined as the mean variance in a wavelength range of 5450\AA\, < $\lambda$ < 5550\AA\, which allowed an accurate S/N estimate free from the strongest absorption or emission lines. A mask was applied such that any single spaxel with mean S/N < 1.5 within the wavelength range was excluded from further binning. This ensured there would be no pollution of useful regions of a cube by particularly noisy pixels. Pixels were infrequently masked with the mean quantity of initially masked pixels per target being 1.4\%, with the most being masked in NGC\,4696 at 6.1\% initially masked, and some targets (e.g. NGC\,1316) requiring only 5 or 6 initially masked pixels.

Following this, the spaxel mask was built upon to include objects not associated with the galaxy, such as foreground stars or bright background objects. A Moffat function \citep{moffat1969theoretical} was applied to the galaxy at V-band which was subtracted from the original V-band slice. The resulting image was scanned using `photoutils' routine included in MPDAF \citep{baconmpdaf2016} with a sigma value defining the threshold for a source to be considered non-continuous with respect to the galaxy light. This threshold varied case by case, depending on aspects such as galaxy brightness, large scale structure, and disky areas of the galaxy. Any segmented regions above the threshold were added to the general mask of pixels to be excluded whilst binning.

\subsubsection{Voronoi Binning}
The MUSE cubes gathered from the sample presented extremely large datasets, with large variations in S/N. The on-source time of each cube varied between objects and surveys but was at a minimum of 590s per spaxel (central region of NGC\,4594) and a maximum of 11400s per spaxel (combined cubes of outer regions of NGC\,4696). To overcome issues associated with low S/N spectra we employed Voronoi binning to ensure the correct treatment of lower S/N areas, whilst maintaining the usefulness of high S/N spectra (usually located in the galaxy core). 

Voronoi binning of S/N (here specifically using the `Vorbin' package of \citealt{Cappellari2003}) gathers data on a 2D spatial plane such that bins reach a minimum summed S/N threshold. Spaxels are optimally arranged into a bin such that the spatial irregularity of a bin is minimised, and simultaneously the sum of the S/N within a bin minimally above the desired S/N threshold (thus high S/N areas are not wasted).

We use a minimum S/N threshold of 100 per \AA\, uniformly for all galaxies in the sample. Due to differences in exposure time, galaxy brightness and other data quality aspects, this results in varying numbers of Voronoi bins per unit area of a given galaxy. The brightest and deepest cubes necessitated many more Voronoi bins per unit area, as many pixels were often left unbinned. Dimmer and shallower targets consist of fewer bins, containing a greater number of binned pixels.

Spectra within a given Voronoi bin were summed, providing a single spectrum representative of the given bin. Final summed spectra were checked to ensure a minimum of S/N=100 was achieved in all bins. This does not however account for spatial co-variance between pixels. Considering the large spatial coverage of the average Voronoi bin (relative to pixel size) spatial correlations are unlikely to impact on derived values in any bin. Furthermore, all Voronoi diagrams are understood to be a general overview of galaxy properties, rather than an exhaustive pixel-perfect map.




\subsubsection{Elliptical Binning}\label{ellip_bin}
When investigating features and trends associated with galactocentric radius, it can be more instructive to bin elliptically. We use elliptical binning to explore features such as ex-situ fraction as a function of radius. This is simply done, with an origin set at the centre of a galaxy. Elliptical rings derived by projected ellipticity are overlayed onto the cube spatial plane with the major axis increasing by defined bin size. Ellipticity and radial increments are derived from 2MASS LGA data, with axes ratios derived from a J+H+K (super) image at the 3-sigma isophote \citep{jarrett20032mass}. This allows all pixels to be binned by galactocentric radius. We performed this using fractions of effective radius (as a standard using 0.1r$_e$ steps) as the increment of major axes, in order to maintain a comparable increment across all galaxies in the sample. The ratio of pixel size to elliptical bin width varies with galaxy size and apparent radius but has a minimum bin width of 11 pixels, occurring in NGC\,1332 due to the elliptical projection as well as the smaller size against the pixel scale. The mean bin pixel width is greatest for NGC\,1316 which has a minimum bin pixel width of 45 pixels.

\subsubsection{Full Spectral Fitting}\label{fsf}
We fit stellar models to binned spectra using the Penalised Pixel-Fitting (pPXF) method \citep{Cappellari2003, cappellari2016improving}. pPXF uses a maximum penalised likelihood approach to extract the stellar kinematics and stellar populations from the spectra of galaxies. Binned spectra are limited to within the wavelength range (4750\,\AA\, < $\lambda$ < 6800\,\AA). This optimally avoids wavelengths that are sensitive to the IMF (this is discussed further in Section \ref{discussion}) as well as avoiding regions particularly affected by sky lines. This wavelength range has been shown to be successful in recovering stellar populations parameters in \cite{guerou2016exploring}.

To ensure the output of pPXF population estimate is physical, we use the regularisation method built into pPXF. Regularisation is intended to smooth over the output weights to provide population estimates that are optimally physical. Weighted templates that have been combined to produce a target spectrum will often be localised to a single nonphysical solution, with many other valid solutions being overlooked, despite their physicality. To produce more representative distributions, regularisation seeks to smooth the solutions to a physical state. The challenge is to smooth the template weights to a solution that most accurately represents observed conditions, whilst not overlooking genuine fluctuations and details present in the model-fit. 

The value of regularisation is adjusted up to a maximum value such that the  reduced $\chi ^2$ of this smoothest solution is no different than 1-$\sigma$ of the unregularised solution \citep{cappellari2016improving}. This has been shown in literature to be an accurate and useful method of galaxy population extraction \cite[see e.g.][]{comeron2015, guerou2016exploring, faifer2017, Ge2019}. 

We use an iterative method to find the most reasonable value of regularisation for each bin in a galaxy cube.  First the noise from the unregularised solution is rescaled so that $\chi^{2}/N = 1$, where N is the number of voxels in the target spectrum.  Next a series of fits using different amounts of regularisation are produced. We record the output $\Delta \chi^2$ for each fit using a particular regularisation, and fit a function to the output  $\Delta \chi^2$ values with respect to the input regularisation guesses. From this function we can find the maximum regularisation parameter which corresponds to $\chi^2 = \chi^2_0+\sqrt{2N}$. We perform this procedure for all bins, obtaining optimal solutions in all cases. The value of the regularisation parameter has little variation for like spaxels of the same galaxy, but can differ between distinct galaxies. Tests were run without regularisation and similar, albeit far noisier, results were obtained.

A total of 552 SSP models, constructed from the MILES spectral library, were used to fit to galaxy spectra \citep{vazdekis2012miuscat}. These models were of Kroupa \citep{kroupa2001variation} revised initial mass function (log slope of 1.3) using BaSTI isochrones, with a metallicity range of -2.27 to +0.4 [M/H] in 12 non-linear steps, and an age range of 0.1 to 14.0\,Gyr in 46 non-linear steps \cite[][]{cassisi2005basti,pietrinferni2006large,falcon2011updated}.

We perform the pPXF routine twice for each bin, once for the robust extraction of stellar kinematics, and a second time for the extraction of stellar populations assuming the fixed kinematics derived during the initial fit. For robust kinematic fitting we use 4th degree additive polynomials. The population analysis uses no additive polynomials, instead using 16th degree multiplicative polynomials and 2nd order regularisation. This has been shown to be effective in kinematic and population analysis \citep[see e.g.][]{guerou2016exploring}.

\subsection{Ex-Situ Fraction Estimation}\label{exfrac}
The mass fractions returned by pPXF in age-metallicity space represent the contribution to the galaxy’s integrated spectrum from different stellar populations. A host galaxy’s chemical evolution will typically progress from older and more metal poor stellar populations, to younger more metal rich stellar populations. The rate of this chemical enrichment is a combination of many complex processes including outflow and recycling of gas, as well as production and removal of heavy elements through stellar evolutionary processes. Many of these processes are dependent on the star formation rate and stellar mass of the host galaxy, and this typically results in lower mass galaxies having lower metallicities (particularly at fixed age). These mass dependent age-metallicity trends have been seen in the Milky Way’s globular cluster system, and those of its satellite galaxies \citep{forbes2010accreted, leaman2013bifurcated, massari2019origin}. 

In \cite{boecker2020galaxy} the authors leveraged these mass dependent chemical evolution trends to identify which regions of the age-metallicity parameter space were likely associated with accreted satellite galaxies of different masses. In that work, the mass fractions in age-metallicity space returned by pPXF’s fit to an integrated spectra were divided into contributions to the light from galaxies of different masses using a simple mass-dependent chemical evolution model. These flexible chemical evolution tracks provided dividing lines in age-metallicity space between galaxies of different masses.

By comparing to mock spectra of EAGLE galaxies where the accretion history was known, \cite{boecker2020galaxy} showed that this approach could largely identify how much mass from low mass satellites the host galaxy had accreted. There are two key limitations of the simple \cite{boecker2020galaxy} models which we improve upon here in order to more robustly compute ex-situ fractions in real galaxies. 

(1) The mass-dependent chemical evolution templates in \cite{boecker2020galaxy} assumed a constant star formation history, simple leaky-box self enrichment, and self-similar internal spreads in metals at fixed age

(2) The association of mass fractions returned by pPXF in age-metallicity space will be non-uniquely attributed to in-situ or ex-situ components of a galaxy when the mass-ratio of a merger is close to unity (and subsequently the chemical evolution pathways are very similar).

The models for estimating ex-situ fractions from full-spectral fitting outputs used here address both of these aspects, and are presented in full in a forthcoming paper (Leaman et al. in prep.). Below we present the relevant ingredients in brief.

\subsubsection{Mass dependent SFH and chemical evolution}
The end goal of these models is to estimate the probabilistic contribution of accreted (‘ex-situ’) material to every age-metallicity bin, for a host galaxy of a given mass. While low mass accretion events ($\mu \leq 1/10$) may separate cleanly in age-metallicity space from the in-situ component of the massive host, more massive mergers will significantly overlap with the ages and metallicities of the progenitor host galaxy. Here we work to quantify this degeneracy and compute the ex-situ contribution for these galaxies when they are dominated in the limiting case by the most massive accretion event. We compute models where the most massive merger has a stellar mass ratio with respect to the present day galaxy of $\mu_{max} = 1/10, 1/5 \text{ or } 1/3$ (e.g., the mass ratios at time of accretion were $\mu_{0} = 1/9, 1/4, 1/2$).

The sole input required for computing the model is an estimate of the present day total stellar mass of the galaxy $M_{*,gal}$. The most massive accretion event is specified in terms of stellar mass as $M_{sat, max}$ = $M_{*,gal}\,\mu_{max}$. The dark matter mass of the host and most massive accreted satellite are assigned stochastically with an empirical stellar to halo mass relation (SHMR) set forth in \cite{leauthaud2012integrated}. This provides an estimate of the likely total virial mass of an observed galaxy or satellite of a given stellar mass. The $1-\sigma$ range of the quoted SHMR fit parameters in that study are used to introduce stochasticity on the adopted DM virial mass. This DM mass for the galaxy is used to specify the likely redshift of infall for the satellite given the DM mass ratio, with functional form and scatter taken from the statistics of subhalo infall in \cite{boylan2011too}. Specifically those authors show there is a mild decrease in the redshift of accretion as the mass ratio increases, which we include to set the accretion redshift when the satellite first enters the virial radius of the host. A coarse prescription for the time for the orbit to decay due to dynamical friction is adopted from Equation 5 of \cite{boylan2011too} with the redshift and mass distribution of orbital circularity taken from \cite{wetzel2010determines}, and is primarily driven by the mass ratio of the host and satellite.

The SFH of the host and most massive satellite are stochastically evolved using empirical relations for galaxy SFRs at different redshifts from \cite{genzel2015combined}. This study characterises the typical specific star formation rate (sSFR) for galaxies of a given mass and redshift. We use this empirical relation recursively starting from redshift zero with the present day mass of the host used to specify the SFR. We proceed back in time in steps of 1 Myr, with the assumption of a constant SFR within those intervals, and subtract off the amount of mass formed in that timestep (e.g., $M_{i-1} = M_{i} - (t_{i}-t_{i-1})SFR|(t_{i})$). A given SFH constructed in this way is obviously approximate, but reproduces the mass dependent evolutionary pathways of galaxies through the star forming main sequence. Each SFH is given correlated noise with a power law slope of $\beta = 2$ to mimic the bursty star formation histories seen in simulated galaxies \citep{iyer2020diversity}. The stochastic assignment of SFHs using both of these aspects typically results in SFRs that vary by $\sim 50\%$ at any given redshift for a single galaxy’s model.

The nominal mass dependent SFHs of the host and most massive satellite are further augmented by a variable quenching time, after which the SF is terminated. In the case of the host, for any given model iteration this quenching time is drawn uniformly between $t_{young} \leq t_{quench} \leq min\left \{ t_{young} + 3.5Gyr, t_{Hubble}\right \}$, where $t_{young}$ is the age of the youngest stellar population detected in the full spectral fit of the observed spectrum. For the satellite galaxy, the quenching time is specified by the time of infall to the host virial radius, plus the orbital decay time as described previously. The 3.5Gyr value acts as a buffer value encompassing uncertainty in the quenching times of galaxies, as well as accounting for uncertainties in t$_{young}$ at high redshift. These models neglect any triggered star formation in the merger, a complex process which may contribute newly formed stars with some metallicity representative of a mixture of the satellite and host gas metallicities at that epoch.

The chemical enrichment in any timestep corresponds to a leaky or accreting box prescription with equal probability. The effective yield is given a mass dependent parameterisation as in \cite{boecker2020galaxy}, with the parameters of the mass scalings chosen stochastically in each model from a range that ensures the final galaxies reproduce the global mass-metallicity relation, and its scatter, observed in SDSS and the Local Volume \citep{gallazzi2005ages, kirby2013universal}. The gas fraction time evolution, which drives any analytic chemical evolution model, is here coupled to the empirical SFHs by re-expressing this quantity in terms of a sSFR ($t_{gas} = (1 + (t_{dep}sSFR)^{-1})^{-1}$, where we adopt the redshift scaling of depletion time from \cite{tacconi2013phibss}. This allows us to self-consistently use the same relations to drive the star formation histories and chemical evolution.

\subsubsection{Model mass fractions and uncertainties}\label{mmr_uncert}
For a single merger history (characterised by the satellite mass ratio and time of accretion) we run 200 iterations of the SFH/chemical model for the host and merging satellite. Due to the stochastic sampling of those model ingredients, the resulting evolutionary tracks in age-metallicity-SFR space span a probable distribution of pathways that a galaxy of that mass might take through this parameter space. These age-metallicity tracks are weighted by the SFR, and the composite density of mass-weights in age-metallicity space are smoothed and binned to the SSP model grid used by pPXF in a metallicity range of -2.27 to +0.4 [M/H] in 12 non-linear steps, and an age range of 0.1 to 14.0 Gyr in 46 non-linear steps.

\begin{figure*} 
	\includegraphics[width=0.9\linewidth]{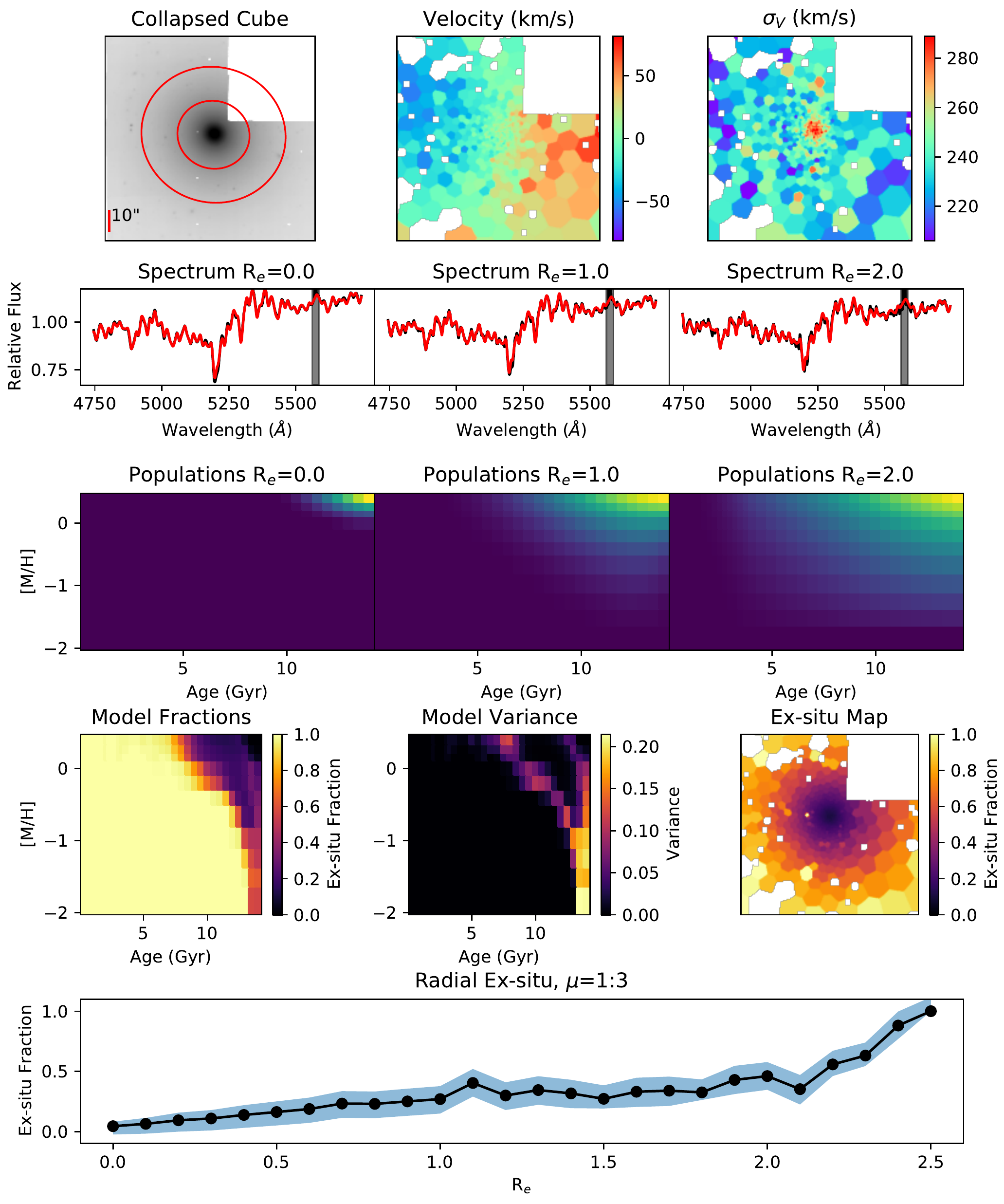}
    \caption{An example process for NGC\,1407. Here we show the analysis and reduction from the collapsed cube, to the final ex-situ estimation. The leftmost panel in the top row shows a full view of the collapsed cube. Colour is inverted, and rings of 1r$_e$ and 2r$_e$ are shown projected onto the cube. The middle panel of the top row shows the stellar velocity as derived by pPXF and applied to a Voronoi binned map of NGC\,1407. The rightmost panel in the top row shows the velocity dispersion for the same map. Other kinematic and chemical properties for NGC\,1407 can be found in Figure \ref{1407_full}. The spectra at 3 different radii are shown in the second row. The r$_e$ value provided is the lower bound of an annulus with an outer boundary of radius (n+0.1)r$_e$. All spectra within these bounds are summed and shown in the panels. The black line represents the original summed spectrum, and the red line shows the final reproduced spectrum derived from full-spectral fitting. A grey shaded region represents a mask used to obscure a strong skyline ([OI]5577\AA). In the third row we show the weighted stellar models in the age-metallicity plane of NGC\,1407 for 3 different annuli in units of effective radius, as derived from the full spectral fitting fit. The left panel of the 4th row shows the ex-situ distribution obtained from the model realisation of many $\mu$=1:3 mergers at all radii, whilst the middle panel shows the resultant variance for all age-metallicity bins in the map. Variance is applied to the total age-metallicity map as per equation \ref{var_eq}. The final Voronoi map of ex-situ stars is shown in the right panel of row 4. This is shown for all galaxies in Figure \ref{ex-sit_vis}. Finally in the 5th row we show the profile of ex-situ fraction in units of r$_e$.}
    \label{1407_poster}
\end{figure*}

\begin{figure*} 
	\includegraphics[width=\linewidth]{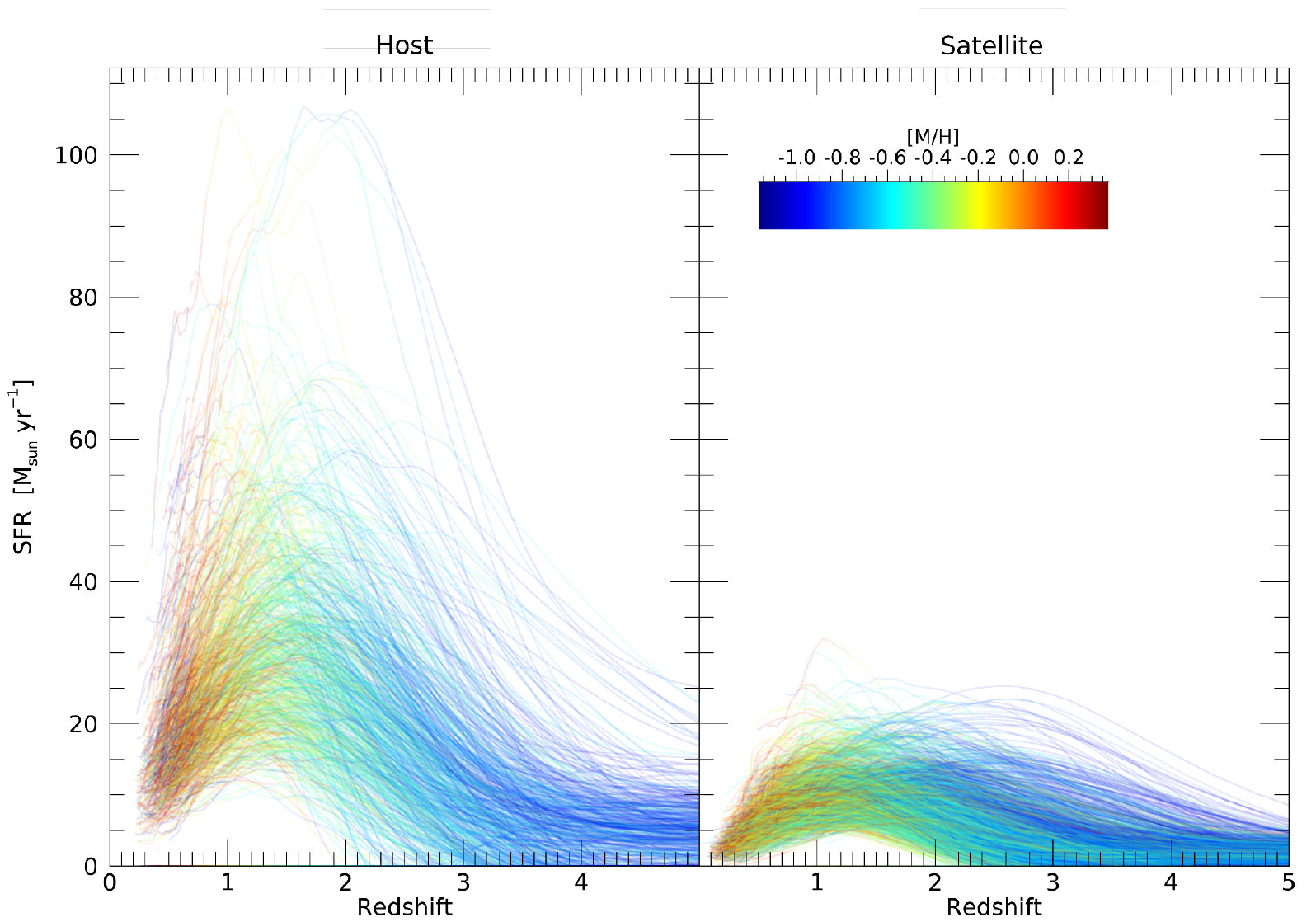}
    \caption{Example empirical SFHs for one merger realisation of NGC\,1380 (merger mass ratio = 1:3). Each thread shows the path in SFR-z space of a single model, with many hundreds of stochastically varied models producing a variance map of expected z=0 population ages and metallicities. Colour shows the mean metallicity increase within the galaxy over time. The time of accretion and quenching time is stochastically generated, informed by the constraints of the youngest stellar population detected in the spectrum (3.5Gyr for NGC\,1380).}
    \label{stochastic}
\end{figure*}

As this is done for the host and the most massive satellite galaxy which merges, for a given cell in age-metallicity space we can ask what is the plausible contribution from overlapping ex-situ/in-situ material. This grid of ex-situ fractions can be multiplied by the mass-fractions returned by pPXF from a full-spectrum fit to an observed galaxy spectrum, to produce an estimate of the contribution of ex-situ material to that galaxy’s spectrum. To quantify the uncertainty in these ex-situ fraction estimates, we run 150 merger histories (stochastically producing different quenching and accretion times) for a model corresponding to a given mass-ratio merger (an example of these stochastic histories is shown in Figure \ref{stochastic}). The variance of these 200x150 models serves as one estimate of the systematic uncertainty in this exercise. The primary uncertainties in the chemical and star formation histories are encompassed within the variations of a single merger history. For these limiting estimates we assume that any cells in the age-metallicity grid which fall below the lower envelope of the stochastic AMR tracks for the host galaxy are from material which must have come from even lower mass satellites - and hence these cells are set to values with $f_{ex-situ} = 1$.  Similarly we assume that any cells with metallicity above the upper envelope of the most massive satellite galaxy stochastic AMRs are contributed to by pure in-situ material, thus accounting for the difficulty in obtaining precise ages in the oldest age bins with any spectroscopic measurement approach. The models are importantly allowing for probabilistic recovery of ex-situ fractions in the the most important regions between these two limits where the AMR distributions
of the host and satellite overlap, which is extremely important for major mergers. This is shown for one example in Figure \ref{1407_poster}.

Final ex-situ fraction and uncertainty can be calculated from a given population as:
\begin{equation}\label{var_eq}
    f_{ex,tot\pm1\sigma}=\frac{\sum_{t,[M/H]}G_{re}\times(0\leq avg(f_{ex})\pm \sqrt{var(f_{ex})}\leq 1)}{\sum_{t,[M/H]}G_{re}}
\end{equation}
Where $G_{re}$ is the product age-metallicity stellar population map derived from full spectral fitting at a given radial annuli.

We stress that while simple, these analytic models are stochastically exploring plausible evolutionary pathways using empirical and physically motivated mass-dependent galaxy scaling relations and theoretical prescriptions. That these models reproduce additional redshift zero scaling relations they were not informed by (scatter and mean of the mass-metallicity relation, SFH-mass relations), is one simple check that they are reasonable ways to characterise the age-metallicity distributions of galaxies. The forthcoming paper (Leaman et al. in prep) will present more details on the model ingredients and calibration checks.

To ensure that assumptions regarding the metallicity gradients of galaxies were valid, extensive investigation into the possible effects of intrinsic metallicity gradients was performed. Appendix A describes this, and highlights the consistency of our method by exploring metallicity gradients with enhanced models and the EAGLE simulations.

\section{Results}\label{results}
The application of model ex-situ fraction grids from tailored simulations allows for the extraction of robust and physical ex-situ fractions for every bin of spectra considered. This method is applied to both elliptical and Voronoi binned galaxy maps, granting insights into galactocentric and more general spatial trends respectively.

\begin{figure*} 
	\includegraphics[width=\linewidth]{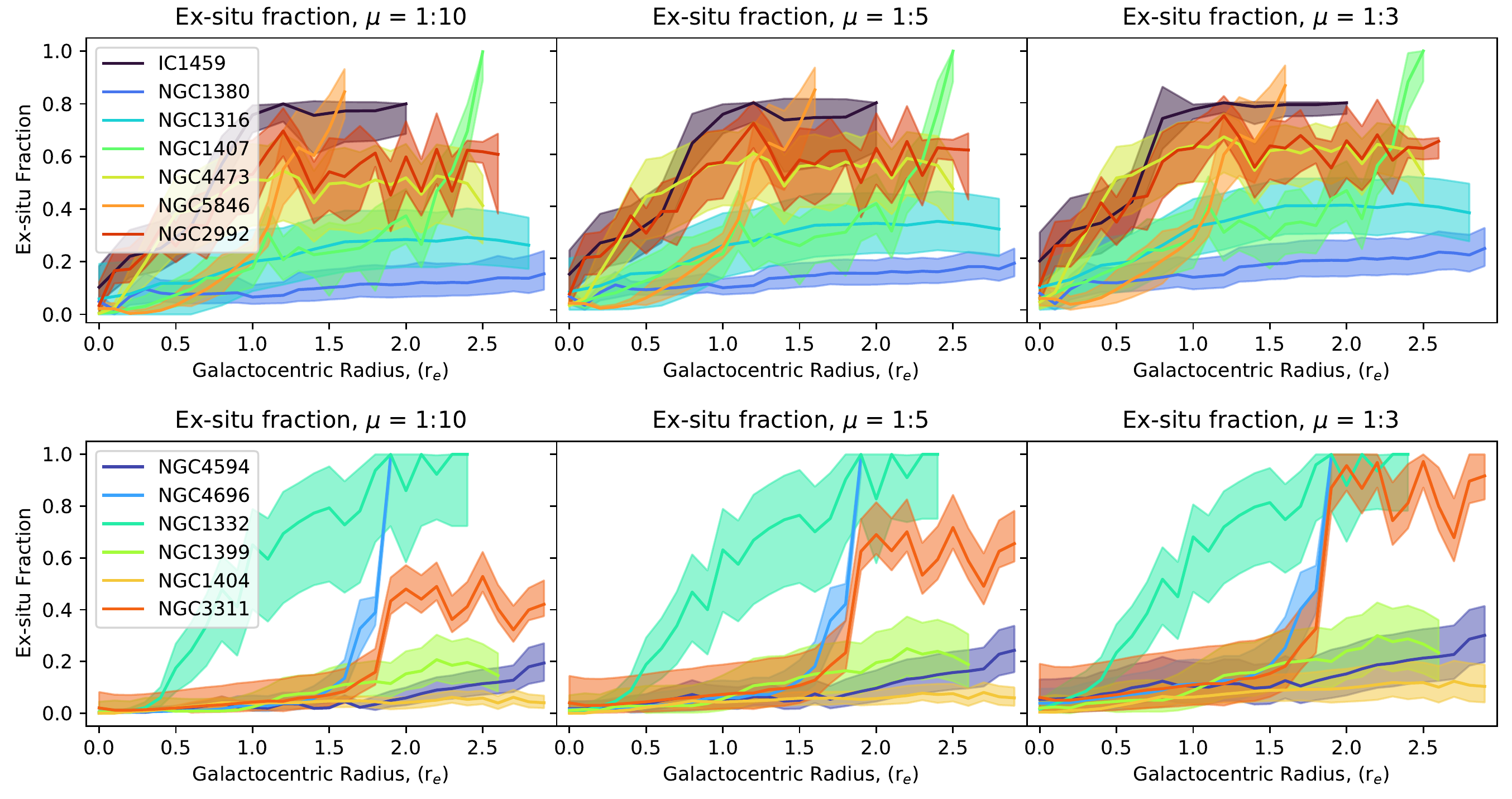}
    \caption{Ex-situ fraction as a function of galactocentric radius for 13 sample galaxies, considering 3 different merger mass ratios ($\mu$) where $\mu$ is the ratio of the present day galaxy mass prior to merger, against the mass of the merging galaxy. A maximum radial extent of 3 effective radii is considered, and a minimum of 1.5 effective radii was a requirement of initial sample definition. Galaxies are split between two sets of figures (upper row and lower row). Division is done purely for ease of legibility and no meaning is ascribed to the split.}
    \label{radius_exsit}
\end{figure*}

A comparison of ex-situ fraction with galactocentric radius for all the sample galaxies is presented in Figure \ref{radius_exsit}. This shows the various ex-situ profiles associated with galaxies, within limited radii, in units of effective radius. This is presented for three different merger mass ratios ($\mu$) where $\mu$ is the ratio of the present day galaxy mass prior to merger, against the mass of the merging galaxy. For each bin, the mean residual value of the ppxf spectrum fit was calculated. Bins with mean residual value of greater than 0.1 (10\% difference from input spectrum) were discarded. In the majority of sample galaxies, no bins were discarded by this condition. 

For all galaxies within the sample, and for all mass merger ratios considered, the ex-situ fraction increases with galactocentric radius. This varies from the barely discernible increase of NGC\,1404 with a maximum of 4.5\% ex-situ fraction at 2.5r$_e$; to IC\,1459 which displays an increase from 10-18\% to 95-98\% ex-situ fraction within 1.5r$_e$ depending on the assumed merger mass ratio. All galaxies contain fewer or no ex-situ stars in the centres ($0R_e < r < 0.25R_e$), with values starting at no more than 18\%, with a mean value of 3.5\% across all sample galaxies and merger mass ratios.

\begin{figure*} 
	\includegraphics[width=\linewidth]{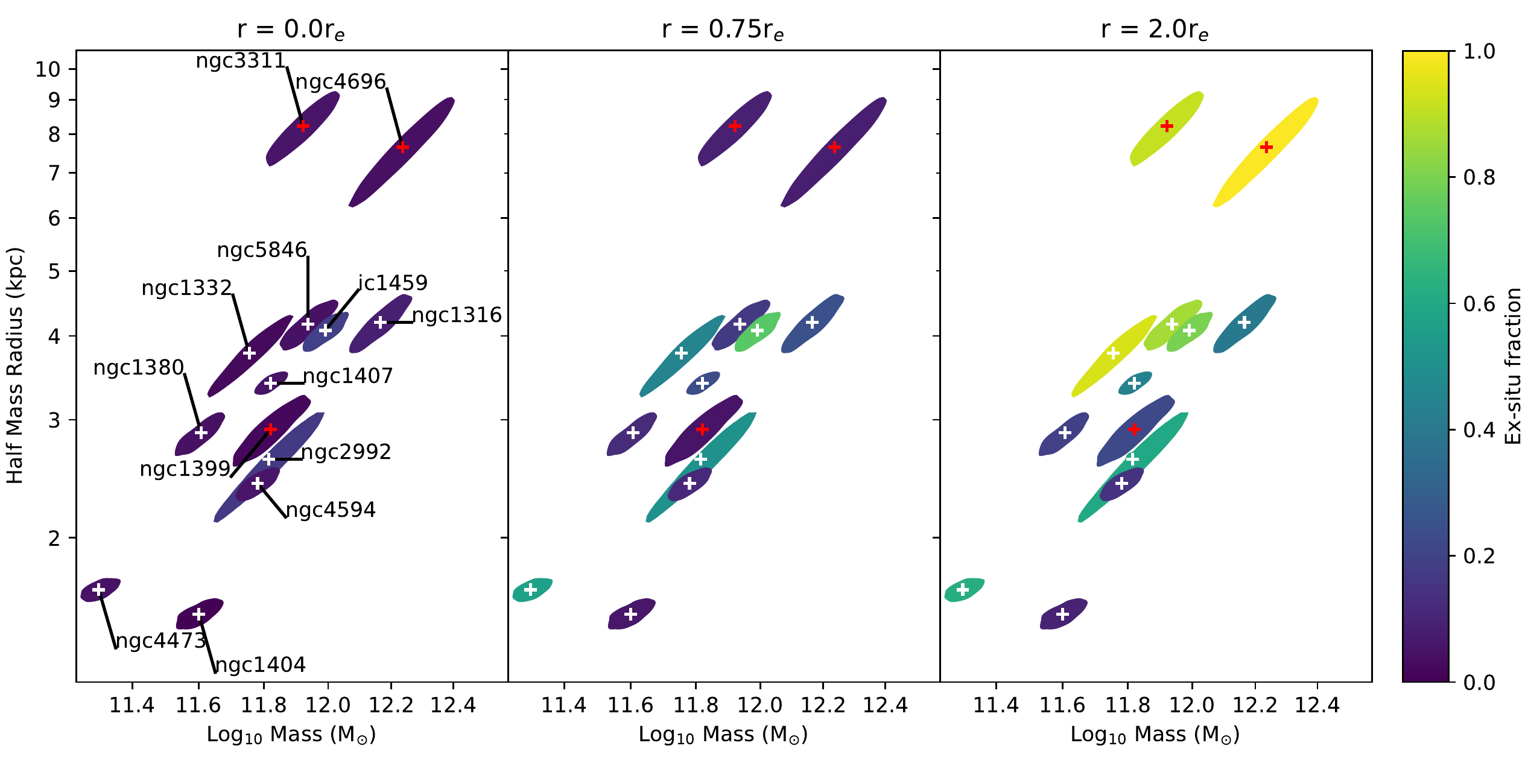}
    \caption{Mass size plane of sample galaxies. Colour shows the ex-situ fraction at various radial intervals for a merger mass ratio of 1:3. Three radial bins are shown, representing the ex-situ fraction found in the elliptical annulus at a given galactocentric radius in units of effective radius. Ex-situ fraction is taken as the mean value within $\pm$0.1re of the specified value. It does not consider ex-situ fraction enclosed within the annulus below the lower bound. Filled ellipses represent the 1$\sigma$ confidence interval for the uncertainty of the galaxy mass and size. Uncertainty in distance affects radius simultaneous to mass-to-light calculations used for total mass estimates, hence uncertainty is simultaneous in the xy plane. Central Dominant galaxies are marked with a red cross. For two galaxies with coverage just short of 2r$_e$, the final ex-situ measurement taken for the third panel is the most extended point possible within cube coverage. For NGC\,4696 this is at 1.8r$_e$, and for NGC\,5846 this is at 1.7r$_e$.}
    \label{ms_col}
\end{figure*}

The differences in ex-situ fraction that increase with radius can be particularly highlighted by taking slices at specific fractions of effective radii. Figure \ref{ms_col} shows the ex-situ fraction for each sample galaxy at 3 different radii. Galaxies are shown as the area enclosing a 1$\sigma$ error of mass and size. The area is coloured according to the ex-situ fraction. Figure \ref{ms_col} clearly shows a trend within the sample of a greater increase in ex-situ fraction with galactocentric radius for galaxies that are both greater in mass, and physical extent at fixed mass. 

\begin{figure*} 
	\includegraphics[width=0.9\linewidth]{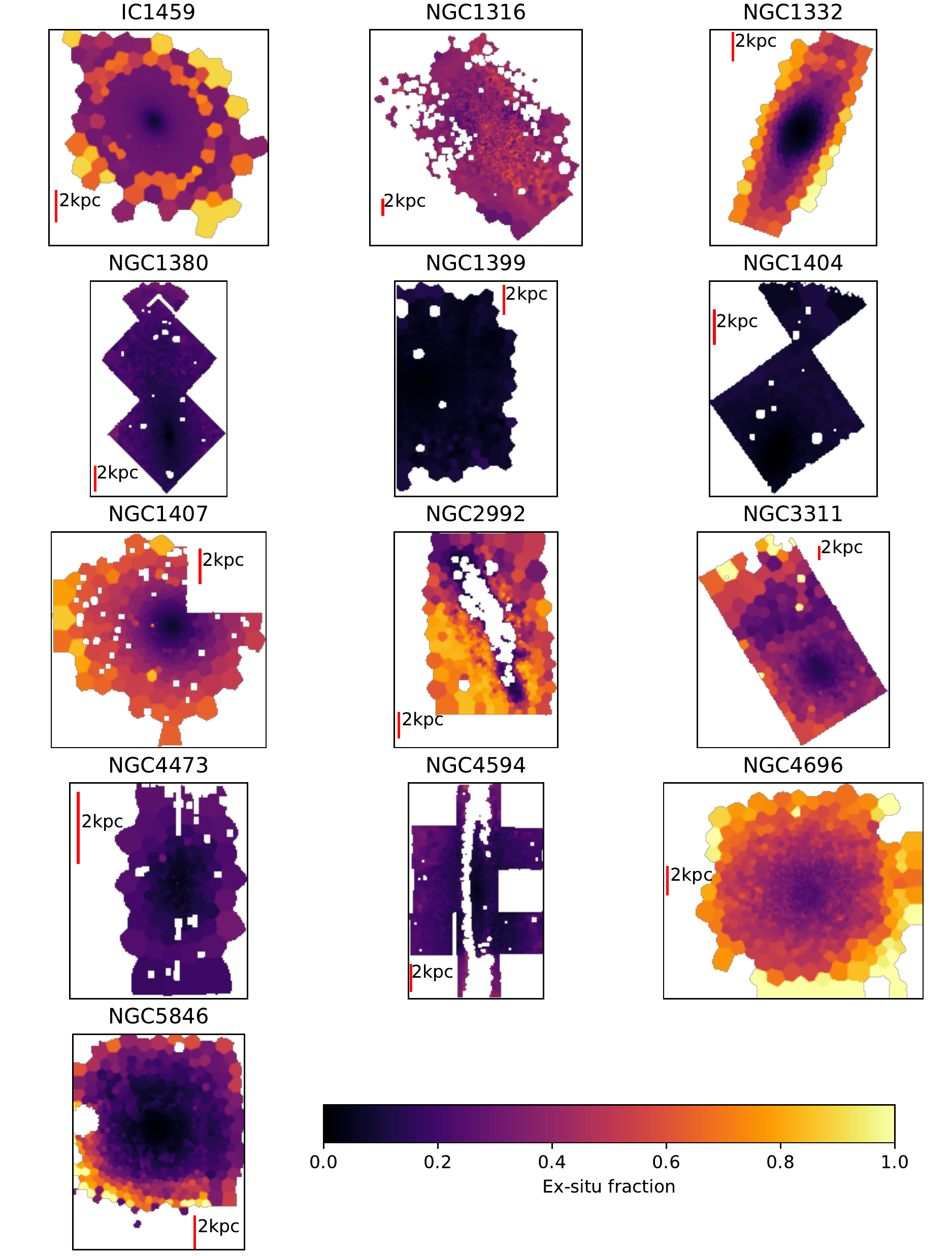}
    \caption{Panels show each galaxy in the sample, with colour showing the ex-situ fraction. Galaxies are Voronoi binned to a signal-to-noise ratio of at least 100 per bin. This Figure shows the calculated ex-situ fraction per bin with an assumed merger mass ratio of 1:3.}
    \label{ex-sit_vis}
\end{figure*}

Estimated Voronoi ex-situ bin fractions are shown in Figure \ref{ex-sit_vis}. This demonstrates the ability for the method to extract ex-situ populations for bins of S/N > 100. For all galaxies, as shown previously, a gradient can be seen in ex-situ fraction in which the fraction of accreted stars increases with radius. For some galaxies, this is closely tied with radius, for example NGC\,1380. This galaxy shows a steady increase in ex-situ fraction at greater radial extents. Other galaxies show both radial and localised increases in ex-situ fraction. For example, NGC\,2992 increases slowly in general ex-situ fraction from the centre, however a feature in the lower-left corner shows an increase in ex-situ fraction. This can be linked to a region of younger than average stellar material that is clearly seen in the lower panels of Figure \ref{2992_full} of the Appendix. 

For galaxies with disks such as NGC\,1332, NGC\,1380, and NGC\,2992, inner regions show stellar material formed primarily of in-situ material, relative to their much more ex-situ dominated outskirts. We stress that low ex-situ fractions do not mean that significant mergers have not occurred. Systems with ex-situ fractions of 20\% could have gained this material in a 1/4 mass ratio merger which can provide significant modifications to the galaxy structure and kinematics.

Sample size prevents any strong conclusions being drawn regarding general galaxy properties, however Figures \ref{radius_exsit} and \ref{ms_col} indicate success in the recovery of ex-situ fraction estimation using population age-metallicity distributions. Below we discuss some aspects of the results with respect to host galaxy properties in this preliminary sample.

\section{Discussion}\label{discussion}

When examining the sample as a whole, and considering the properties of the galaxies involved, a number of striking features are revealed. For the 50\% least massive and least extended galaxies (those with a half-mass radius $<$3kpc), ex-situ fraction remains low throughout, with a mean ex-situ fraction at 2 effective radii of 22\% for all merger mass ratios. For the most massive and most extended galaxies (those with a effective radius $>$3kpc), ex-situ fractions increase more drastically, with a mean ex-situ fraction at 2 effective radii of more than 58\% for the same merger mass ratios.

\begin{figure*} 
	\includegraphics[width=\linewidth]{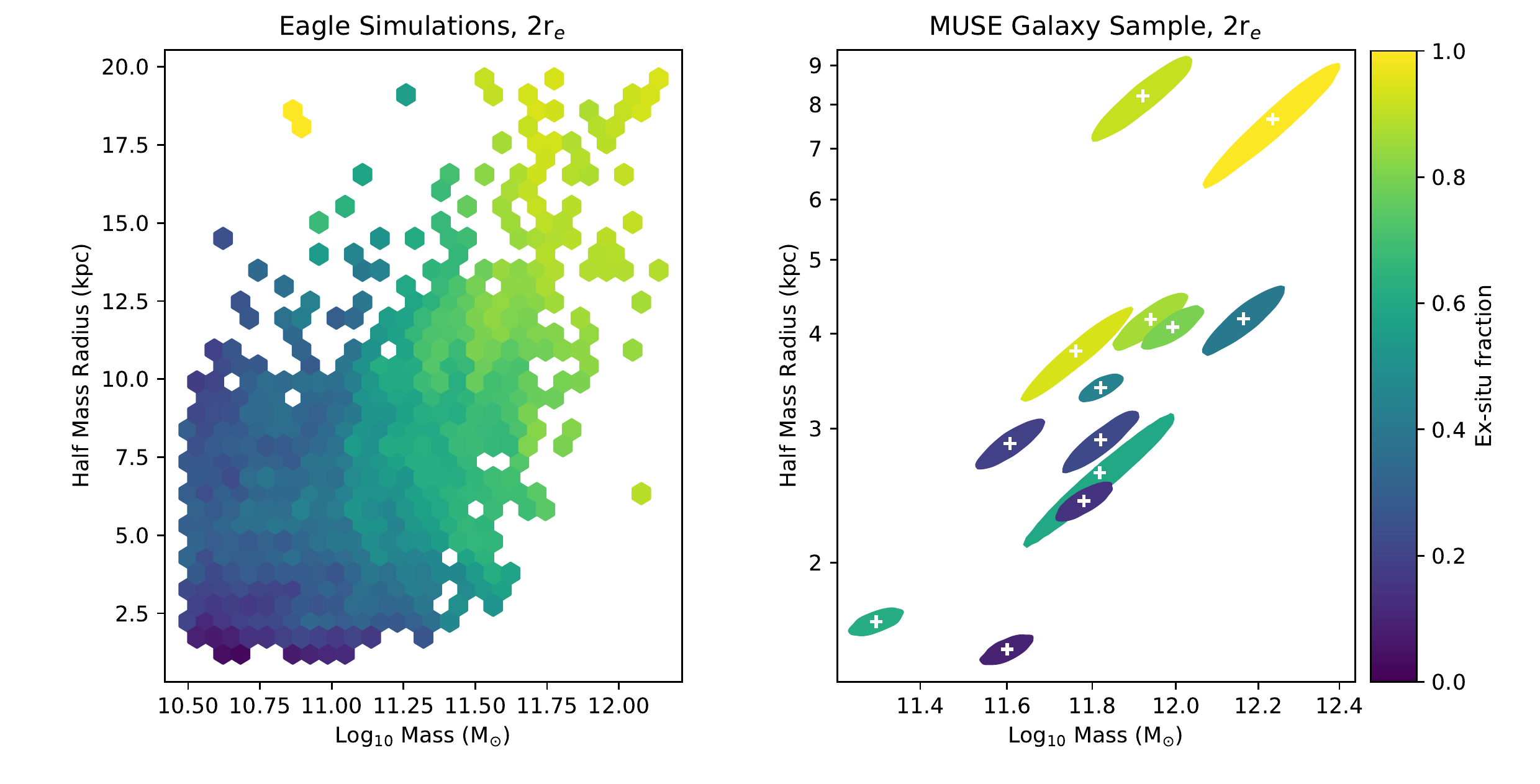}
    \caption{A side-by-side comparison of the ex-situ fraction in galaxies from the Ref-L0100N1504 EAGLE simulation, and MUSE galaxies from this paper. The left panel shows 3643 binned EAGLE galaxies, with each bin showing the mean ex-situ fraction of galaxies between 1.9 < r$_e$ < 2.1. Only galaxies with a minimum of 500 stellar particles were considered. The right panel shows the es-situ fraction of MUSE galaxies between 1.9 < r$_e$ < 2.1. EAGLE galaxies tend to be slightly more extended at fixed mass than real galaxies.} 
    \label{eagle_comparison}
\end{figure*}

When viewing galaxies at fixed stellar mass there is a general trend that extended objects experience a greater increase in ex-situ fraction with radius. This is in line with analytic expectations and is seen in the EAGLE simulations \citep{davison2020eagle}. We directly compare to analysis of the EAGLE simulations in Figure \ref{eagle_comparison}, using data from \cite{davison2020eagle}. Here we see side by side comparisons of the ex-situ fractions of stars found in galaxies at 2r$_e$ with respect to the mass-size plane. We see very similar ranges and differential trends in ex-situ fraction with mass and size between the MUSE analysis of this paper, and galaxies from the Ref-L0100N1504 EAGLE simulation \citep{schaye2015eagle, crain2015eagle}. The Ref-L0100N1504 simulation is a periodic volume, 100cMpc (co-moving kiloparsecs) on a side, realised with 1504$^3$ dark matter particles and an initially equal number of gas particles. Stellar particles are traced throughout the simulation providing in- or ex-situ tags for each star present in a given galaxy at z=0. The limits of mass and size analysed in this paper are indicated on the EAGLE panel as a dashed box. It should be noted that Ref-L0100N1504 EAGLE galaxies tend to be slightly more extended at fixed mass than real galaxies. This has a number of potential explanations. One is the result of over-efficient feedback within the EAGLE simulations \cite[see e.g.][]{crain2015eagle} which moves galaxies of fixed M$_*$ into haloes of higher M$_{200}$ potentially extending the galaxies. A second explanation is described in \cite{ludlow2019energy}, where 2-body scattering can artificially inflate galaxies. Regardless of any mass-size offsets, both the theoretical and observational datasets show a clear preference for higher ex-situ fractions in larger and more massive objects. Galaxy density appears to be correlative factor with ex-situ fraction and the gradient of accreted stars with galactocentric radius.

In comparison to \cite{oyarzun2019signatures} who use stellar metallicity profiles to estimate the ex-situ fractions of MaNGA galaxies, results are reasonably well matched. The results in the \cite{oyarzun2019signatures} similarly show an increase in ex-situ fraction with both mass and galactocentric radius, however the scatter in these values at $>2r_e$ essentially cover the entire ex-situ axis. Similar to population analyses in \cite{edwards2020clocking} we find population gradients in essentially all the galaxies studied, occurring strongly with both age and metallicity. These gradients are indicative of low-metallicity young stars accreting to the outskirts of the galaxies in this sample. Similarly to \cite{rodriguez2016stellar} we find that in the case of major mergers (assumed in \cite{rodriguez2016stellar} to be $\mu > 1/4$) the contribution to the ex-situ population is significant with gradients that change the ex-situ fraction by as much as 90\% within 3re. In all cases where we assume $\mu > 1/3$ the ex-situ fraction recovered from the sample galaxies was largest compared to other assumed merger masses.

Final mass-size uncertainties were calculated using a Monte-Carlo approach. Distance and projected size uncertainties impacted on the area with which to calculate mass using mass to light ratios. As such, uncertainties can be seen as a misshapen ellipse enclosing the 1$\sigma$ mass-size probability, either diagonal in the mass-size plane (distance uncertainty dominated), or more horizontal in the mass axis (mass-to-light ratio uncertainty dominated). Galaxies with more precise distance estimates (such as NGC\,1404) exhibit relatively low mass-size uncertainty. Values of mass and size and associated uncertainties can be found in Table \ref{gal_vals}. When considering all mass ratios, the average uncertainty in ex-situ fraction at any radius was $\pm$8.4\%.

Three of the galaxies included in the sample are considered to be central dominant galaxies (cD) according to literature classifications. All other galaxies in the sample are variously field or satellite galaxies. The three galaxies with cD classification are NGC\,1399 (Fornax), NGC\,4696 (Centaurus), and NGC\,3311 (Hydra) and are marked with a red cross in Figure \ref{ms_col}. It is perhaps surprising to find that NGC\,1399 seems to contain little ex-situ stellar material in the region imaged. This may have to do with the ill-defined and estimated radius for cD galaxies. For such multi-component galaxies, r$_e$ is a simplification, and as such classifying parameters in units of r$_e$ can be misleading. Light on the outskirts of cD galaxies extends well into the intracluster medium, and the point at which galaxy light transitions to intracluster light is poorly defined if measurable \citep{seigar2007intracluster}. As such, ex-situ estimates for these cD galaxies represent a lower limit of the possible values. Most likely (especially in the case of NGC\,1399) the true radius is much larger than normal methods have provided here, and we only probe the very inner regions of NGC\,1399. Further limitations arise in the age of NGC\,1399 and the age of its mergers. Due to poor age resolution in current spectral models, mass accreted onto NGC\,1399 in early times could potentially be missed, and therefore much ex-situ mass is misidentified as in-situ mass. Estimates from \cite{spavone2017vegas} put the total accreted fraction (within $\approx$8r$_e$) at a more reasonable 84.4\%. On the other hand it is entirely possible that this is an accurate view of NGC\,1399. This is a statistically poor sample and we expect a scatter in the true distributions of galaxy ex-situ fraction with galaxy type, mass, and radius. Larger statistically significant samples are necessary to judge this with any certainty.

In the cases of NGC\,4696 and NGC\,3311 the radial profile of ex-situ fraction increases dramatically over ~0.2-0.6r$_e$. Both galaxies show similar extents and are the largest galaxies by area in the sample. The rate of increase in ex-situ fraction was examined against various galaxy properties such as morphological type and inclination with no convincing correlation found. One possible reason for the increase is the result of dust in the centres of these galaxies. Both the aforementioned galaxies display irregular dusty features in the centre. Potentially this dust skews the age estimate of the full-spectral fitting to older populations, causing a lower fraction of the stars to be classified as ex-situ. Once the elliptical binning moves outside of this dusty range, the ex-situ estimate returns to its unaffected state, and gives the impression of a rapid rise in ex-situ fractions. This appears visible in the very few galaxies present with diffuse dusty features. Further tests are needed to help correct for the impact of dust on stellar population properties. Further potential impacts on radial ex-situ profiles by intrinsic gradients are discussed in Appendix A.

Stellar population models were obtained from the MILES spectral library \citep{vazdekis2012miuscat}. The choice of stellar population models can affect the derived weighted grid, slightly shifting the values of age and metallicity according SSP differences. Much literature has discussed the differences encountered in using alternative SSP models \citep{conroy2009propagation, conroy2009propagation2, conroy2010propagation2, fernandes2010testing, delgado2010testing, baldwin2018comparison, martins2019testing, knowles2019stellar}. Despite the differences in available SSP models, testing with different models would be unlikely to change results in a differential sense with any significance \citep[see also][]{boecker2020recovering}. Though alternative SSP models would possibly give shifted absolute age and metallicity values, in relative terms the populations would remain distinct and separable. 

A further parameter that we did not vary was the assumed IMF, which is unlikely to be exactly Kroupa-like for Early-Type galaxies \citep{vaughan2018stellar, smith2020evidence}. The implications of a variable IMF were particularly explored in \cite{clauwens2016implications}. At fixed age an SSP with bottom heavy IMF will have more low mass stars and therefore the age recovered with a Milky Way-like IMF would be older. The wavelength range used for the MUSE analysis in this paper avoids some of the most IMF sensitive regions of the spectrum such as the NaI doublet \citep{schiavon1997near} and CaI \citep{smith2012stellar}.

While our approach can only provide approximate separation of the in- and ex-situ components, the benefits of our data and models are two-fold over past approaches. First, the use of spectroscopy allows for recovery of ex-situ components within the centres of the galaxies, where accretion is still known to deposit ex-situ material. Secondly, the physically and empirically motivated models used to decompose the recovered age-metallicity distribution account statistically for the expected overlap in structure and stellar populations in massive accretion events, which are the most transformative to the host galaxies.

We aim to extend this work further over larger and more representative mass ranges. With complete IFU coverage of galaxies out to 2 effective radii we can endeavour to examine ex-situ fractions of galaxies at every mass range. Better understanding of the influence of environment, as well as cross-validation of our estimates of ex-situ fractions with those of complementary approaches (such as from GC population analysis and photometric imaging of the faintest outer halos \citealt[e.g.][]{rejkuba2014tracing}) will allow for a useful census of accreted material in local galaxies. 

\begin{table*}
\caption{Sample galaxy values used to estimate mass and size. Mass and size values are the mean value from Monte-Carlo simulation considering all available sources of error. Magnitudes and galaxy type were obtained from the 2Mass Large Galaxy Atlas \citep{skrutskie2006two}, whilst distance was taken from various sources:}
\label{gal_vals}
\begin{tabular}{llrrrrrrrlll}
Galaxy & Type & \multicolumn{1}{l}{Ks-band} & \multicolumn{1}{l}{$\pm$} & \multicolumn{1}{l}{J-K} & \multicolumn{1}{l}{$\pm$} & \multicolumn{1}{l}{Distance} & \multicolumn{1}{l}{$\pm$} & \multicolumn{1}{l}{{[}Mass{]}} & M$_{\sigma}$ & r$_e$ & r$_{e,\sigma}$ \\
 &  & \multicolumn{1}{l}{mag} & \multicolumn{1}{l}{} & \multicolumn{1}{l}{} & \multicolumn{1}{l}{} & \multicolumn{1}{l}{Mpc} & \multicolumn{1}{l}{} & \multicolumn{1}{l}{M$_{\odot}$} & M$_{\odot}$ & kpc & kpc \\ \hline
IC1459 & E3;AGN & 6.928 & 0.016 & 0.904 & 0.03 & 28.7 & 1.80 & 11.99 & 0.07 & 4.08 & 0.25 \\
NGC1316 & SA(rs)b;Sy1;2 & 5.694 & 0.016 & 0.877 & 0.03 & 20.8 & 2.00 & 12.16 & 0.10 & 4.18 & 0.41 \\
NGC1332 & SAb & 7.155 & 0.016 & 0.949 & 0.03 & 22.9 & 3.40 & 11.76 & 0.14 & 3.79 & 0.56 \\
NGC1380 & S(s)0 & 6.971 & 0.016 & 0.911 & 0.03 & 18.6 & 1.40 & 11.61 & 0.08 & 2.86 & 0.21 \\
NGC1399 & SA0 & 6.431 & 0.016 & 0.924 & 0.03 & 18.2 & 1.75 & 11.82 & 0.10 & 2.90 & 0.28 \\
NGC1404 & cD;E1;pec & 6.941 & 0.016 & 0.948 & 0.03 & 17.1 & 0.76 & 11.60 & 0.09 & 1.54 & 0.07 \\
NGC1407 & E1 & 6.81 & 0.016 & 0.952 & 0.03 & 20.6 & 0.97 & 11.82 & 0.06 & 3.40 & 0.16 \\
NGC2992 & E0 & 8.714 & 0.016 & 1.123 & 0.03 & 38.0 & 7.70 & 11.82 & 0.19 & 2.63 & 0.53 \\
NGC3311 & Sa\_pec;Sy1 & 8.154 & 0.016 & 0.918 & 0.029 & 45.7 & 5.35 & 11.92 & 0.11 & 8.22 & 0.98 \\
NGC4473 & SA(s)a;Sy1.9 & 7.287 & 0.016 & 0.906 & 0.03 & 15.2 & 0.57 & 11.30 & 0.07 & 1.67 & 0.06 \\
NGC4594 & S0- & 5.04 & 0.016 & 0.933 & 0.03 & 8.99 & 0.52 & 11.78 & 0.08 & 2.41 & 0.14 \\
NGC4696 & cD;E+2 & 7.192 & 0.016 & 0.972 & 0.03 & 38.9 & 6.59 & 12.23 & 0.16 & 7.64 & 1.33 \\
NGC5846 & E7/S0 & 7.044 & 0.016 & 0.953 & 0.03 & 26.3 & 2.10 & 11.94 & 0.08 & 4.17 & 0.33
\end{tabular}
\end{table*}

\section{Conclusions}\label{conclusion}
Thirteen MUSE targeted galaxies were analysed to examine the ex-situ populations within galaxies across a small mass-limited area of the mass-size plane.  Galaxies were chosen for their large coverage and spatial resolution, allowing for a well resolved examination of ex-situ populations with galactocentric radius. By combining analytic models with weighted populations extracted by full spectral fitting we were able to distinguish ex-situ stellar material that was incongruous to in-situ stellar matter. In all galaxies examined, ex-situ fraction increases with radius. This increase was shown to be particularly strong in more massive galaxies, and more extended galaxies. Whilst less extended and less massive galaxies experienced radial ex-situ fraction increases up to 5-40\% at 2$R_e$, the more extended and massive galaxies in the sample show radial increases up to 25-100\%. The result that the magnitude of radial ex-situ fraction increase seems to correlate with both mass and physical galaxy extent agrees well with existing simulation predictions \cite[see e.g.][]{davison2020eagle}.

These estimations for the lower-limit of galaxy ex-situ fractions agree with theoretical predictions that expect accreted material from the lowest mass (and hence lowest metallicity) merged galaxies to preferentially remain in the outskirts of galaxies. This reveals how coverage to a minimum of two effective radii is necessary for exploration of galaxy populations, in order to avoid considering only unmixed in-situ portions of massive galaxies.

By estimating and mapping ex-situ stars in a sample of galaxies, we demonstrate the power in tracing accreted populations with IFU spectroscopy. In conjunction with other methods to quantify accreted stars, we now endeavour to utilise the full power of spectroscopically derived full star formation histories. With estimates of the distributions in stellar age and metallicity, we can expect to better constrain and understand galaxy assembly history.
\section{Acknowledgements}
This research has made use of the services of the ESO Science Archive Facility. Based on observations collected at the European Southern Observatory under ESO programmes: 094.B-0298 (P.I. Walcher, C.J), 296.B-5054 (P.I. Sarzi, M), 0103.A-0447 \& 097.A-0366 (P.I. Hamer, S), 60.A-9303 (SV), 094.B-0321 P.I. (Marconi, A), 094.B-0711 (P.I. Arnaboldi, Magda). Some contextual panel views were based on observations made with the NASA/ESA Hubble Space Telescope, and obtained from the Hubble Legacy Archive, which is a collaboration between the Space Telescope Science Institute (STScI/NASA), the Space Telescope European Coordinating Facility (ST-ECF/ESA) and the Canadian Astronomy Data Centre (CADC/NRC/CSA).  This work was completed with support from the ESO Studentship Programme, and the Moses Holden Studentship. GvdV acknowledges funding from the European Research Council (ERC) under the European Union’s Horizon 2020 research and innovation programme under grant agreement no. 724857 (Consolidator Grant ArcheoDyn). AB acknowledges funding by the Deutsche Forschungsgemeinschaft (DFG, German Research Foundation) -- Project-ID 138713538 -- SFB 881 (``The Milky Way System'', subproject B08). This research made use of Astropy, \url{http://www.astropy.org} a community-developed core Python package for Astronomy \citep{astropy2013, astropy2018}.

\section{Data Availability}
The data underlying this article are available in the ESO science archive facility (\url{http://archive.eso.org/cms.html}).


\bibliographystyle{mnras}
\bibliography{biblio} 


\appendix
\setcounter{figure}{6}
\subsection*{Appendix A: Verification of Metallicity Effects}
\subsubsection*{A.1 Eagle Simulation Verification}
The methodology used to ascertain the ex-situ fractions of sample galaxies relied on an expectation of metallicity evolution with respect to ex-situ fractions, and largely ignored possible radial metallicity gradients attributed to internal mechanisms independent of the ex-situ fractions. The true interplay between in- and ex-situ populations that result in the final galaxy metallicity gradient are hugely complex and difficult to disentangle. We therefore present robust tests regarding this possible metallicity complication to ensure that our final results are valid irrespective of small underlying internal processes.

We use 718 galaxies from the EAGLE simulations \citep{crain2015eagle,schaye2014eagle} to assess the validity of the metallicity assumptions. Galaxies were selected to be of `red type' (more spheroidal and passively evolving), by selecting only galaxies with co-rotational values of $\kappa _{co} < 0.4$ \citep{correa2017relation}. Galaxies were further chosen to ensure a minimum of 500 stellar particles were present within 3re, and that star formation in the prior 1Gyr remained low. Each galaxy was subsequently divided into five radial bins between 0 and 5 effective radii.

For each radial bin of all galaxies, mock spectra were produced by summing weighted spectra from an AMR grid derived from stellar particles present in a galaxy radial bin. Gaussian noise was applied and spectra were convolved with Gaussian functions to replicate a spectral line width consistent with 100km/s velocity dispersion. These mock spectra were then analysed with pPXF to extract an estimated AMR. This allowed us to also ensure that degeneracy was not affecting our final results.

As it was not feasible to run simulations for each individual galaxy, ex-situ fraction was estimated using the method described in \cite{boecker2020galaxy}. As described in Section \ref{exfrac}, mass dependent chemical evolution trends are used to identify which regions of the age-metallicity parameter space are likely associated with ex-situ populations. A total of 500 Monte-Carlo simulations were produced for each bin of every galaxy, providing a robust estimate to the recovered ex-situ fraction.

This recovered estimate was then measured against the true value from the EAGLE galaxies. The mean absolute difference in ex-situ estimate varied for mass bins. In the lowest and highest mass bins, ex-situ fraction was either very low or very high respectively. This allowed easy recovery of ex-situ fraction. For intermediate masses, the AMR was more complex and showed greater variation between galaxies. At the very most, the mean difference between the recovered and the true ex-situ fraction was 9.5 percentage points. Even if all galaxies in our main MUSE sample were incorrect by 9.5 percentage points, no results would be significantly affected, and a division between the ex-situ fractions of the lowest and most massive galaxies would remain.

\begin{figure} 
	\includegraphics[width=\linewidth]{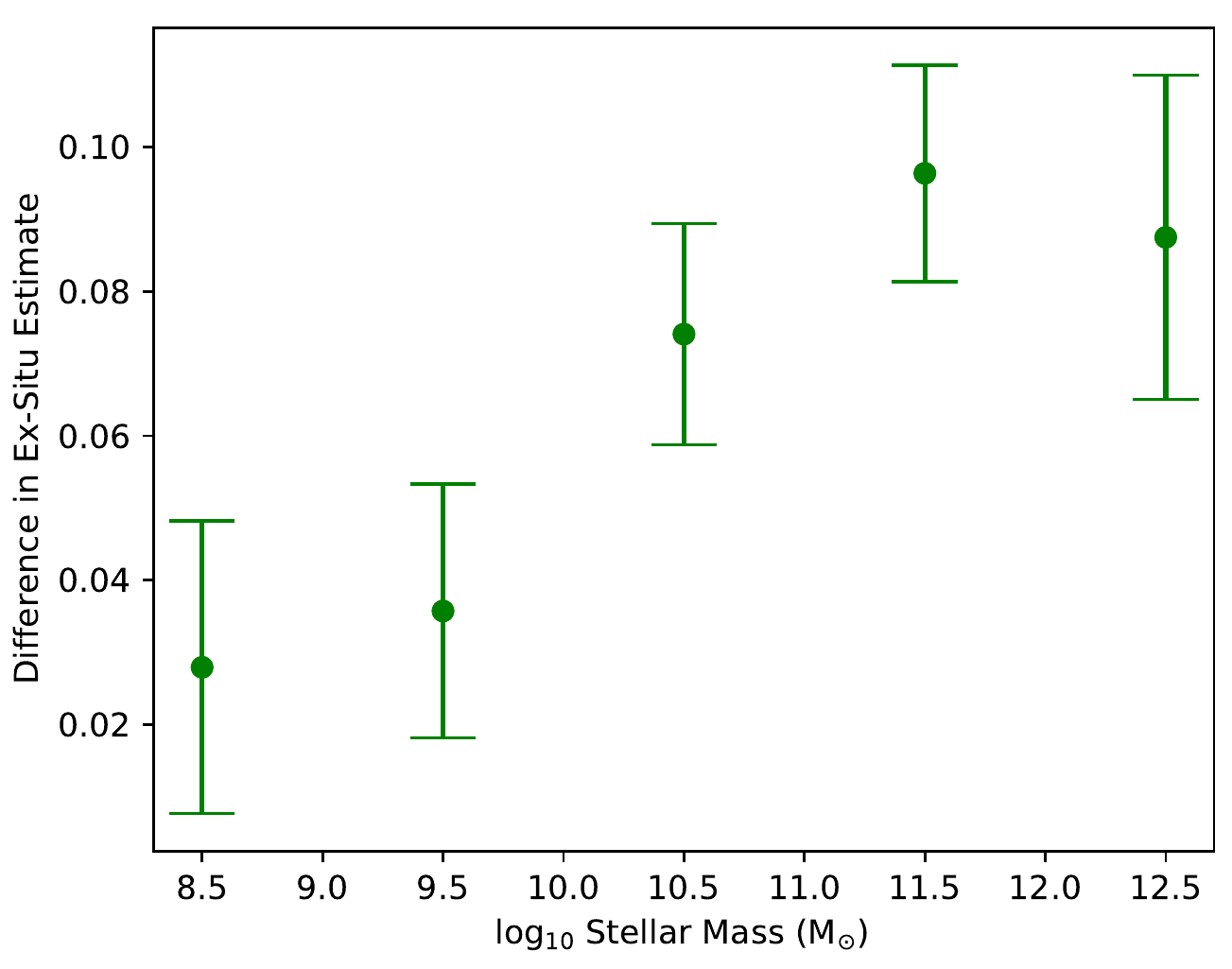}
    \caption{Mean absolute difference of ex-situ recovery from real, shown for five stellar mass bins. A total of 718 EAGLE simulation galaxies were used, with mock spectra generated for each in 5 radial bins between 0 and 5 effective radii. The y-axis shows the value of difference between the EAGLE galaxies and the real galaxies studied in this paper. Green circles indicate the mean difference in ex-situ fraction across all radial bins for all galaxies, for 5 different stellar mass bins. Error bars indicate the standard deviation in difference within the mass bin. Points are shown at the centre of their mass bin, and cover $\pm$0.5 dex in stellar mass.}
    \label{subsel}
\end{figure}

\subsubsection*{A.2 Adaptive Metallicity Gradient Modelling}
While galaxies show a wide diversity of radial metallicity gradients, all observations to date quantify average metallicity from the \emph{total} stellar population light at any location (e.g., an unknown blend of in-situ and ex-situ stars). The interplay between the radial variation in ex-situ fraction that we recover, and any intrinsic radial metallicity gradients in the in-situ (and ex-situ) stars are therefore complex. We explore changes in recovery of ex-situ fractions when assuming intrinsic radial metallicity gradients, by adapting the models used as per Section \ref{mmr_uncert}. To examine possible effects in detail we have constructed a version of our analytic ex-situ fraction model grids which include spatial variations in the chemical evolution. We describe below how we use recent simulation and observational results to guide our choices in exploring any potential impacts of radial metallicity gradients of the in-situ and ex-situ stellar components separately.

We proceed by computing our model chemical evolution tracks for the in-situ and ex-situ stellar populations as described in Section \ref{mmr_uncert}. For a given stochastic model of many merger histories and chemical evolution histories, these produce density distributions for the ex-situ and in-situ components within the AMR grid, and these are used to decompose the returned pPXF mass fractions into ex-situ fractions. Here we simply modify this exercise to produce model density distributions of the ex-situ and in-situ components at each measured radius (0.1 R/Re steps) within the galaxy, and modify the mean metallicity of the AMR density distributions according to a particular radial metallicity gradient.

For the in-situ component we adopt a logarithmic metallicity gradient of d[M/H]/dRe = -0.2 dex/Re, which \cite{zhuang2019dynamical} showed is typical for galaxies of similar mass to NGC\,1380. For the ex-situ component we adopt three possible cases:  i) no radial metallicity gradient, ii) the same radial metallicity gradient as the in-situ component and iii) a metallicity gradient found for only the ex-situ particles in galaxies of that mass using TNG-50 simulations (\cite{nelson2019first}, Zhuang, Leaman, Pillepich, in prepaparation). 

\begin{figure} 
	\includegraphics[width=\linewidth]{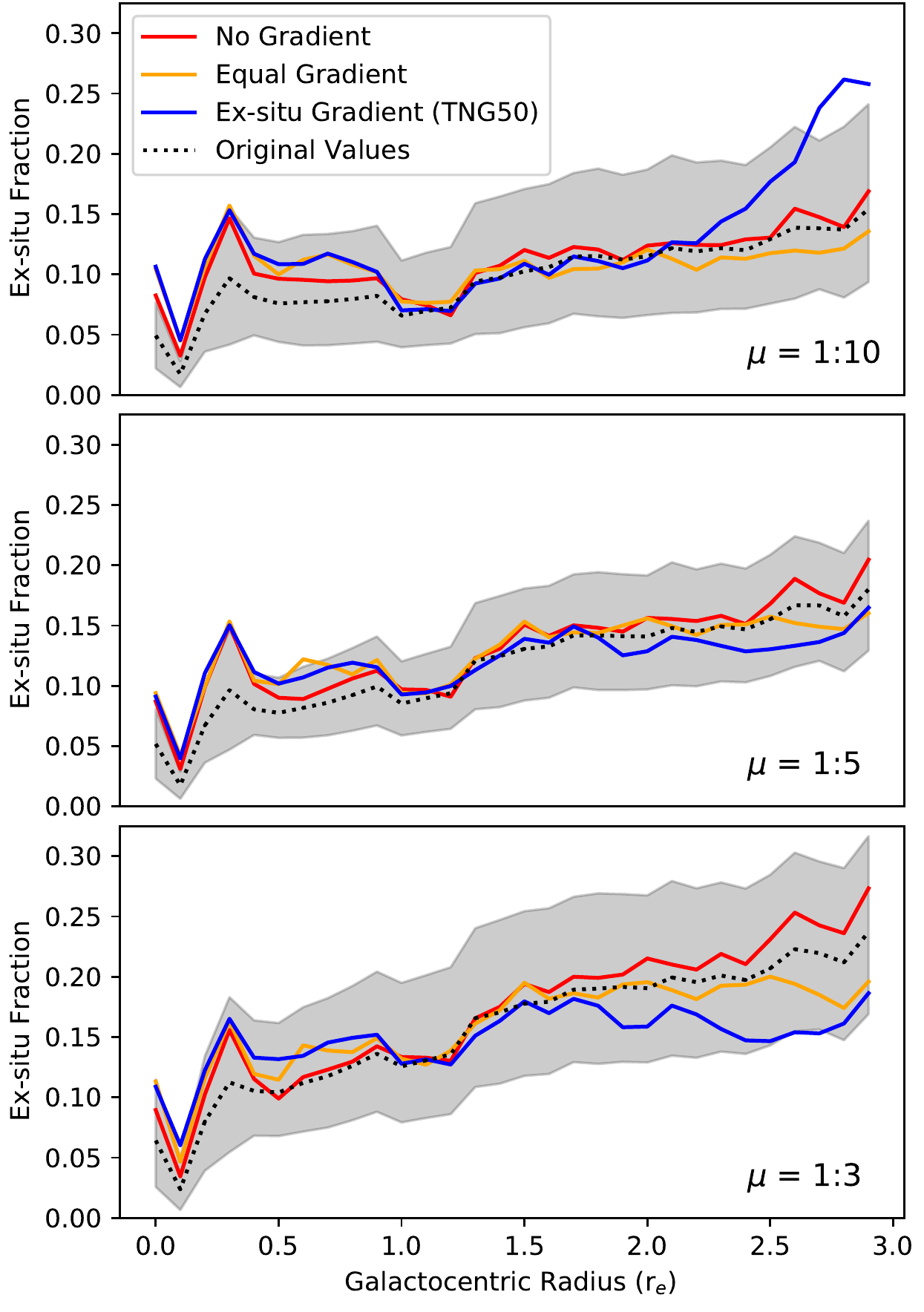}
    \caption{Ex-situ fraction profile estimations for NGC\,1380, using three different possible intrinsic metallicity gradients. The chemical evolution models were adapted according to three intrinsic metallicity cases, and their predicted ex-situ fractions were computed as described in Section \ref{mmr_uncert}. The three cases were: no radial metallicity gradient (red line); equal in-situ and ex-situ gradient magnitudes (orange line); ex-situ metallicity gradient as predicted by the TNG-50 simulations (blue line). These are overplotted on the predicted ex-situ fractions from the base models with no extra consideration of intrinsic profiles (black dotted line), with uncertainties shown with a grey hilighted region.}
    \label{comp_mods}
\end{figure}

In each case, the final model ex-situ fraction grid was used to assign the mass density in the AMR plane (derived from pPXF) an ex-situ fraction. However this was done at every radius with a locally specified ex-situ grid tailored to that radius in each of the three cases for the radial metallicity gradients described above. The attached figure shows how these new model grids (which incorporate reasonable amounts of radial dependence on metallicity informed from observations and simulations), modify the ex-situ fraction profiles for NGC\,1380.  It is evident that the changes introduced by this are important, but within the systematic uncertainties due to degeneracies, regularisation, and resolution metallicity and ages of populations. Figure \ref{comp_mods} shows the results of the adapted models, with coloured lines indicating the newly derived ex-situ profiles for NGC\,1380. All are found to be largely within the intrinsic uncertainty, as demonstrated by the grey shaded regions.

\subsection*{Appendix B: Galaxy Sample - Interesting Cases and Points of Note}

The galaxy sample was sub-selected from the wider available MUSE data as described in Section \ref{method}. Figure \ref{subsel} shows this selection from available MUSE data.
\begin{figure} 
	\includegraphics[width=\linewidth]{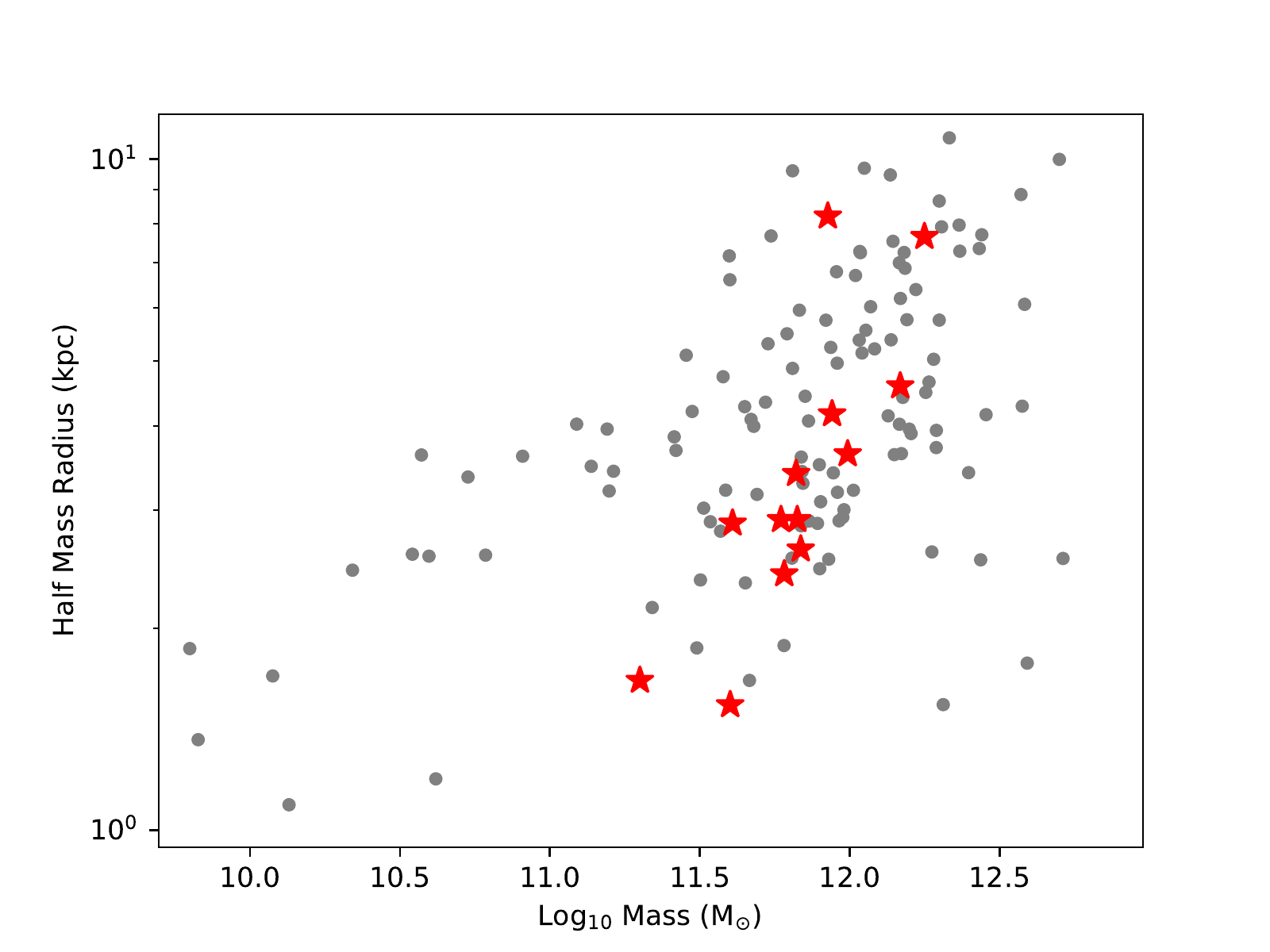}
    \caption{Distribution of available MUSE targeted galaxies (grey dots), shown against final sample galaxies (red star) after sub-selection described in section \ref{method}. Distance for non-selected galaxies was calculated with a simplistic recessional velocity calculation assuming Hubble flow and a H0 value of 67km/s/Mpc.}
    \label{subsel}
\end{figure}
In this section we will lay out some points of interest and note regarding particular sample galaxies.
\subsubsection*{IC1459}\label{1459}
In Figure \ref{1459_full} we see a clear counter-rotating core at the centre of IC\,1459. This is well known and well studied \citep{franx1988counterrotating, cappellari2002counterrotating, prichard2019unravelling}. Counter-rotation has long been associated with galaxy mergers \citep[see e.g.][]{bertola1988counter}. The galaxy appears to show nested rings of distinct populations which are particularly visible in the M/H panel of Figure \ref{1459_full}. This feature was questioned considering the lack of kinematic features, however the ring can clearly be seen as a region of gas experiencing starformation, for instance in H-Alpha or in H-Beta, shown in Figure \ref{1459_hbeta} of the Appendix. 
\begin{figure} 
	\includegraphics[width=\linewidth]{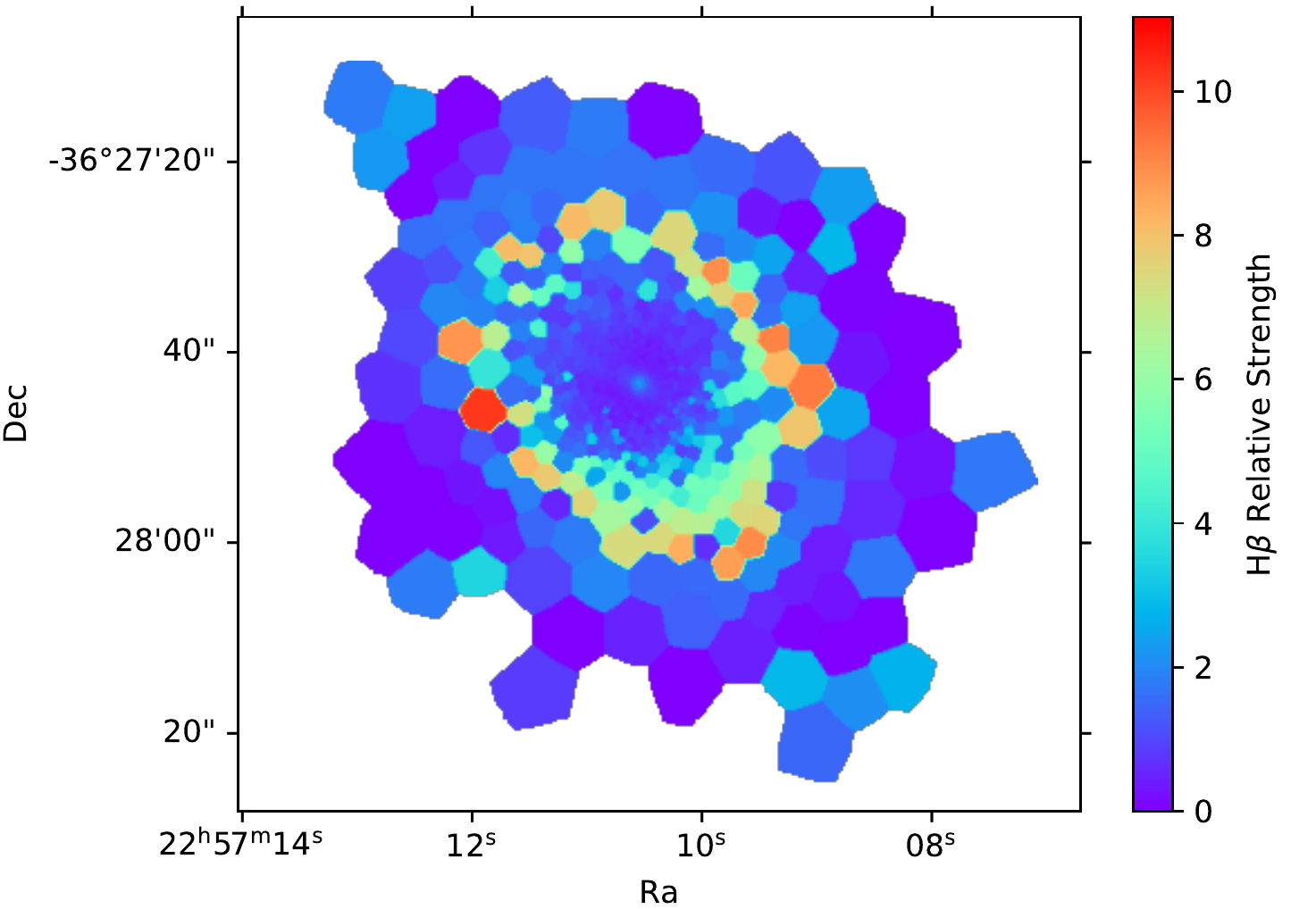}
    \caption{H-Beta in IC\,1459, showing a clear ring and matching to that seen distinctly in metallicity. Colour shows relative magnitude of H-Beta.}
    \label{1459_hbeta}
\end{figure}
\subsubsection*{NGC\,1316}
NGC\,1316 is a giant elliptical galaxy in the Fornax cluster, and one of the brightest radio sources present in the sky \citep{geldzahler1984radio}. The general galaxy morphology was extensively studied in \cite{schweizer1980optical, schweizer1981optical}. \cite{mackie1998evolution} use X-ray data to provide strong evidence from \cite{schweizer1981optical} that the galaxy underwent an intermediate merger, further evidenced by the presence of shells and strong tidal features, and evidence of accretion \citep{horellou2001atomic, serra2019neutral}. 
\subsubsection*{NGC\,1404}
The elliptical galaxy NGC\,1404 is located in Fornax and is one of the brighter members. NGC\,1404 is known to have historical and ongoing interaction with the Fornax central galaxy NGC\,1399, with a recent fly-by at around 1.1-1.3Gyr ago \citep{sheardown2018recent}. The interactions between NGC\,1404 and NGC\,1399 appear to have resulted in the stripping of globular clusters from NGC\,1404 which are now found to be associated with NGC\,1399 \citep{Bekki2003}. MUSE observations of the galaxy resulted in a chance nova outburst observation seen in data taken in November 2017 \citep{smith2020discovering}. In Figure \ref{1404_full} we find a double sigma spike in the centre of NGC\,1404. This matches to a small area of counter-rotation also in the centre, which is the most likely cause of this double peak in dispersion.
\subsubsection*{NGC\,2992}
NGC\,2992 is a constituent of Arp\,245, along with NGC\,2993 and the dwarf galaxy A245N. NGC\,2992 appears to be in the early stages of interaction \citep{guolo2021exploring}. IR imaging shows a very weak stellar stream connecting NGC\,2992 with NGC\,2993. We see the affects of this stream in Figures \ref{2992_full} and \ref{ex-sit_vis}. In Figure \ref{2992_full} we see younger than average populations in the lower left of the 3-6Gyr stellar age panel in the same area as the connecting stellar stream. These younger stars dilute older populations as seen in the 12-14Gyr panel. Furthermore in Figure \ref{ex-sit_vis} we see higher than average ex-situ fractions in the lower left of the panel, in the direction of the stellar stream.
\subsubsection*{NGC\,4594}
NGC\,4594, otherwise known as `The Sombrero Galaxy' is an elliptical galaxy with an inclination of 84 degrees. Its prominent disk has made it a target of both scientific and artistic value. A large tidal stream of stellar material has recently been identified in \cite{martinezdelgado2021feather}, further pointing to evidence of historical accretion. The dust lane presents particular challenges in full spectral fitting. Though pPXF has advanced tools that are well equipped to deal with dust features in galaxies, the thick dust lanes in NGC\,4594 appears to be too thick for pPXF to accurately handle. Residuals from the spectral fitting were on average 30\% higher in the thick dusty regions, compared to residuals in the bulge. This pushed many bins above the threshold residual limit and they were excluded. This also causes issues (to a far lesser degree) in Figures of NGC\,1316 and NGC\,2992.
\subsubsection*{NGC\,4696}
NGC\,4696 is the central galaxy of the Centaurus cluster of galaxies. A clear asymmetric dust feature reaches out, visually disrupting the otherwise seemingly placid galaxy. There are also less visually distinct filaments of gas and dust seen in line-emission studies \citep{fabian1982optical, laine2003hubble, sparks1989imaging}. \cite{crawford2005extended} suggest the filaments are the result of buoyant gas bubbles lifting the gas from the galaxy centre, and find areas of low X-ray pressure supporting this hypothesis. Others suggest the gas features are evidence of the accretion of a gas rich galaxy \citep{farage2010optical}.
\subsubsection*{NGC\,5846}
NGC\,5846 is a massive elliptical galaxy with a broadly spherical shape, with an axis ratio of $\sim$0.95 \citep{jarrett20032mass}. The galaxy is well known to be a near-by example of an AGN galaxy, with extensive studies in X-ray \citep{trinchieri2002peculiar, Machacek_2005}. NGC\,5846 is well studied largely due to both its proximity as an AGN galaxy, and its extensive globular cluster count \citep{forbes1996hubble}. There is strong evidence that NGC\,5846 experienced a recent flyby by its visual companion, NGC\,5846A. Indications hint at tidal stripping of gas and stars from a number of close companions in the associated galaxy group \citep{mahdavi2005ngc, zhu2016discrete}

\begin{landscape}
\begin{figure}
	\includegraphics[width=0.96\linewidth]{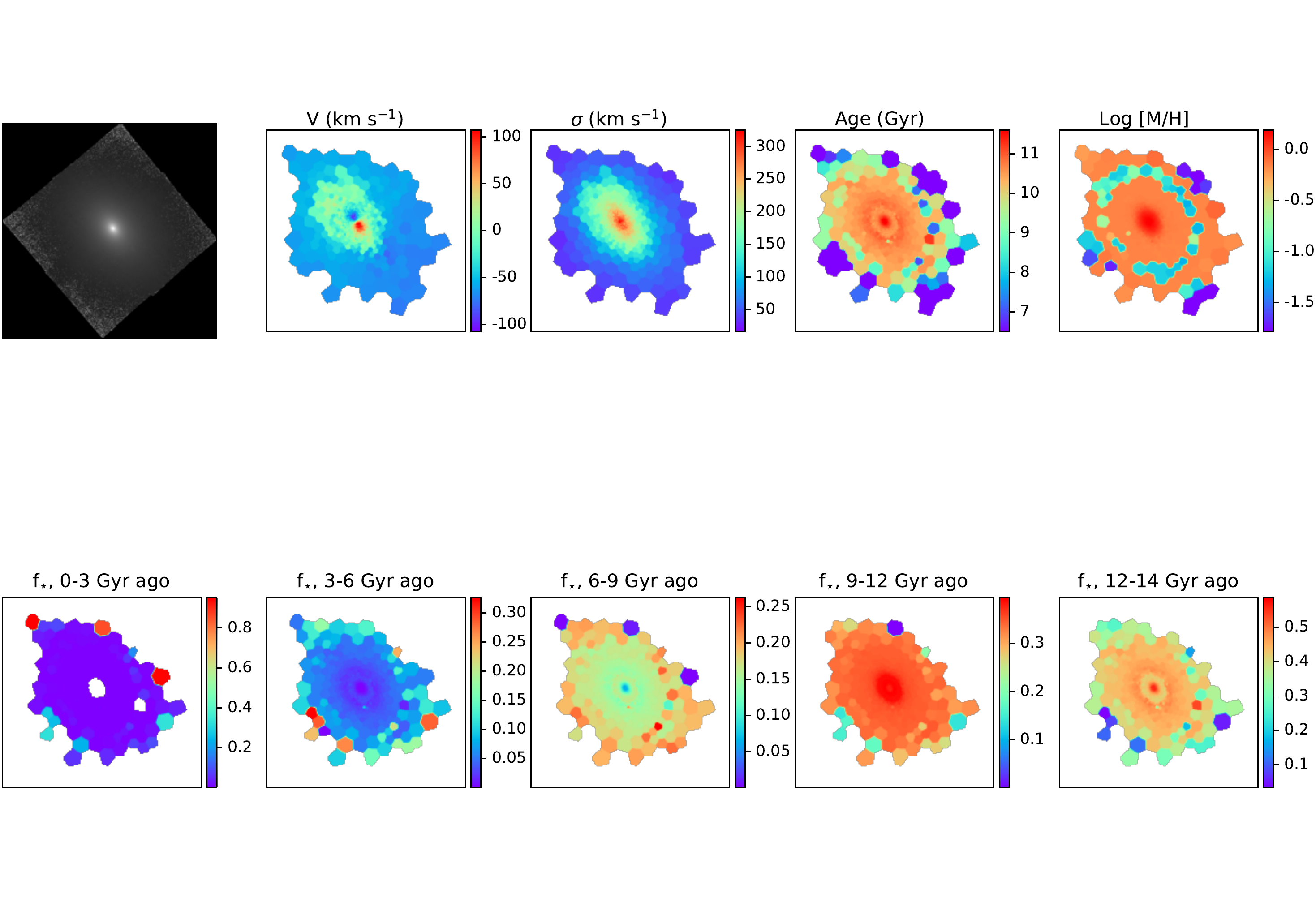}
    \caption{IC\,1459. Top left panel shows the collapsed MUSE cube. Other top row panels show stellar parameters of velocity, velocity dispersion, age, and metallicity. Lower panels show stars selected by age. Colour shows the fraction of stars in a bin that are of a specific age}
    \label{1459_full}
\end{figure}
\end{landscape}
\begin{landscape}
\begin{figure}
	\includegraphics[width=0.96\linewidth]{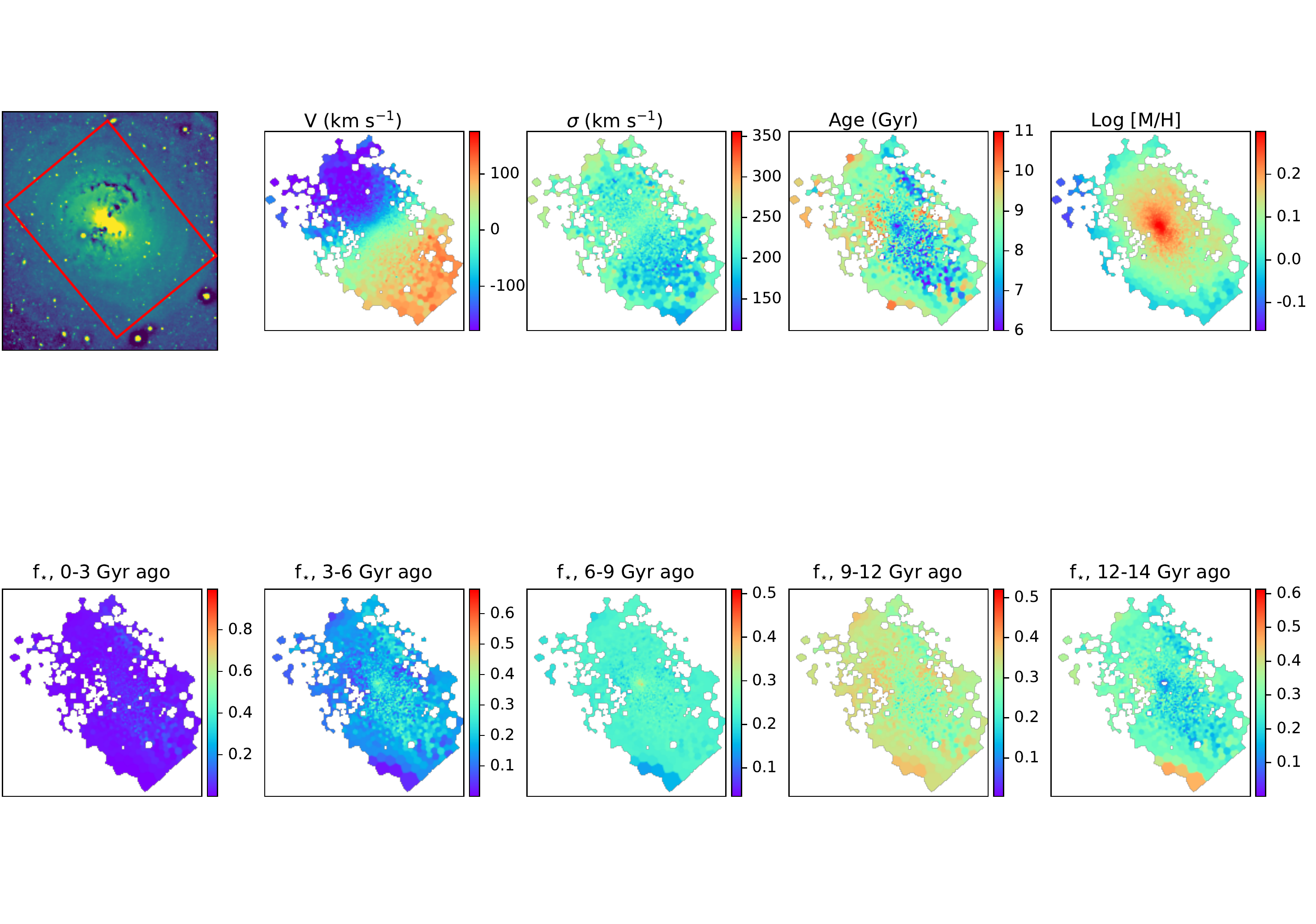}
    \caption{NGC1316. Top left panel shows a Hubble view, with the MUSE cube footprint overlayed in red. Unsharp mask has been applied to highlight galaxy shells (unsharp amount: 25, radius: 25). Other top row panels show stellar parameters of velocity, velocity dispersion, age, and metallicity. Lower panels show stars selected by age. Colour shows the fraction of stars in a bin that are of a specific age.}
    \label{1316_full}
\end{figure}
\end{landscape}
\begin{landscape}
\begin{figure}
	\includegraphics[width=0.96\linewidth]{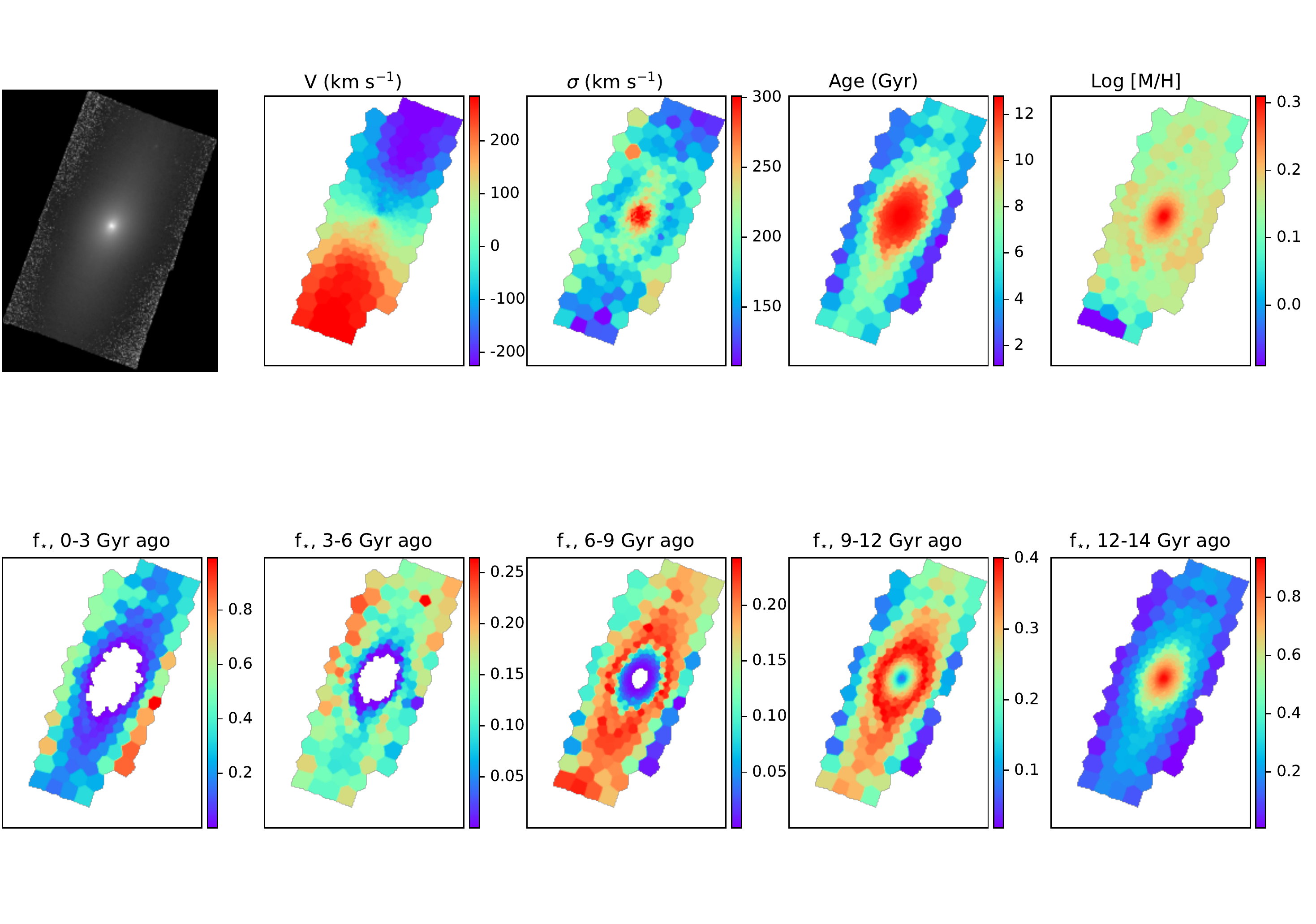}
    \caption{NGC\,1332. Top left panel shows the collapsed MUSE cube. Other top row panels show stellar parameters of velocity, velocity dispersion, age, and metallicity. Lower panels show stars selected by age. Colour shows the fraction of stars in a bin that are of a specific age}
    \label{1332_full}
\end{figure}
\end{landscape}
\begin{landscape}
\begin{figure}
	\includegraphics[width=0.96\linewidth]{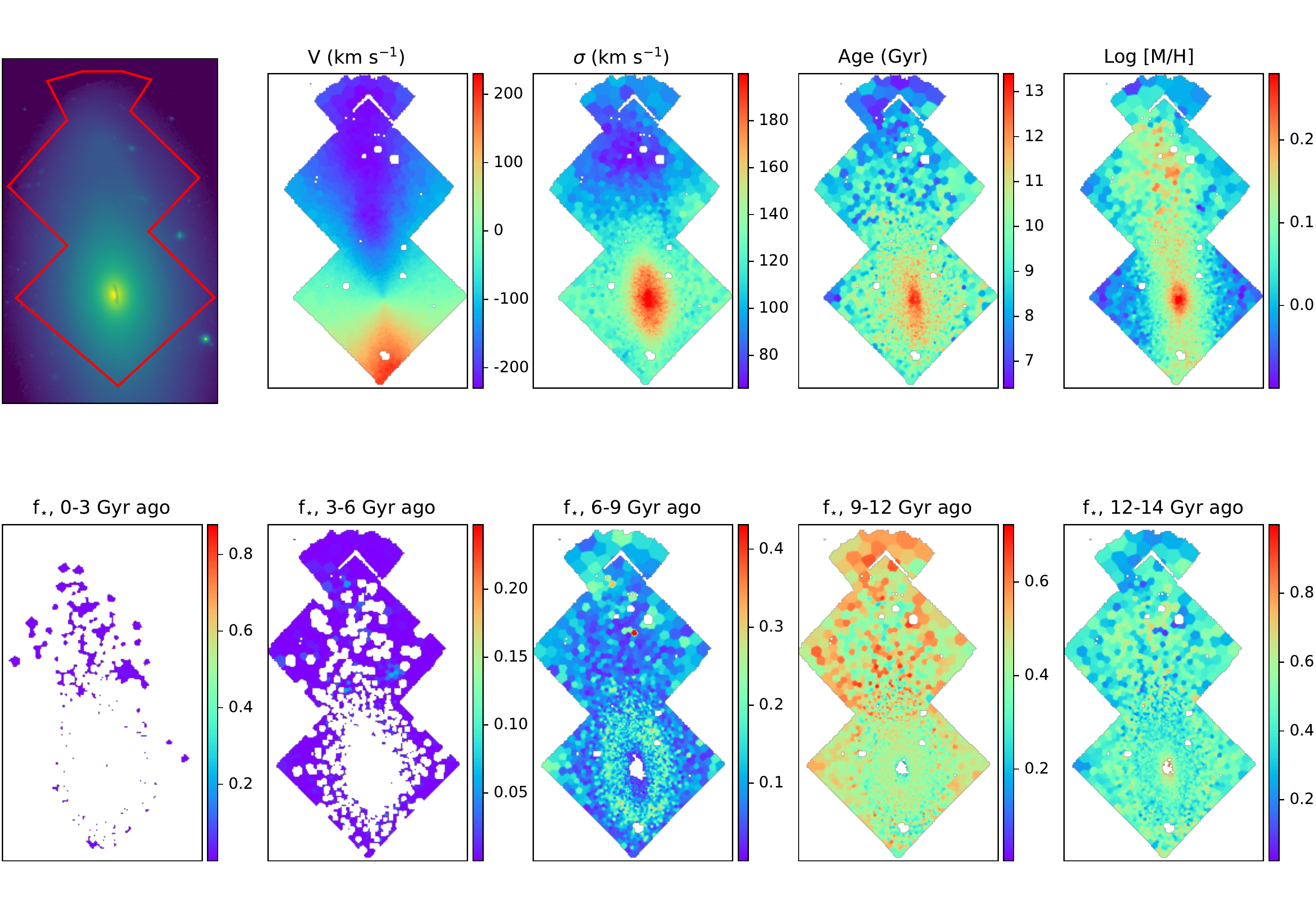}
    \caption{NGC1380. Top left panel shows a Hubble view, with the MUSE cube footprint overlayed in red. Other top row panels show stellar parameters of velocity, velocity dispersion, age, and metallicity. Lower panels show stars selected by age. Colour shows the fraction of stars in a bin that are of a specific age.}
    \label{1380_full}
\end{figure}
\end{landscape}
\begin{landscape}
\begin{figure}
	\includegraphics[width=0.96\linewidth]{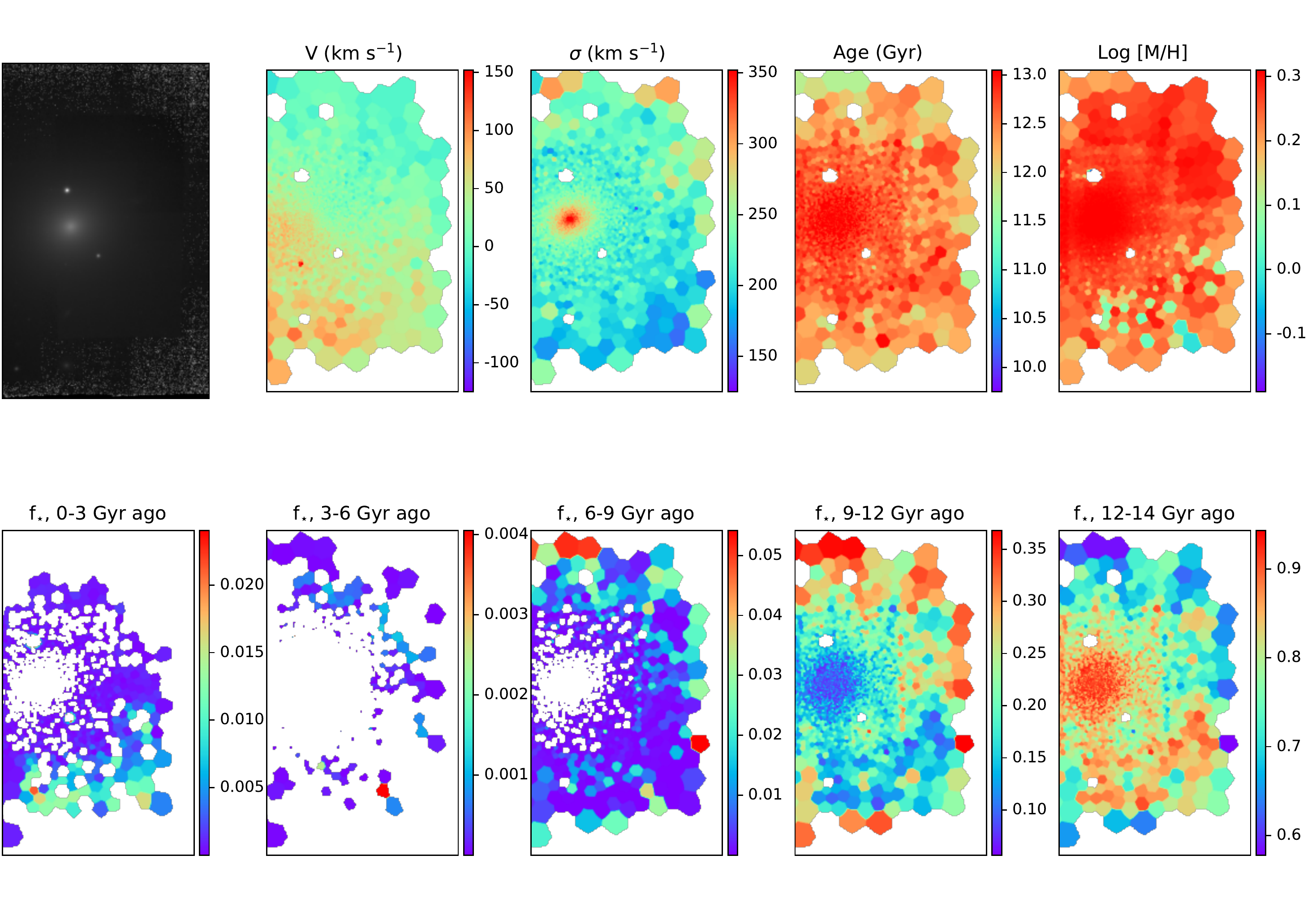}
    \caption{NGC1399. Top left panel shows the collapsed MUSE cube. Other top row panels show stellar parameters of velocity, velocity dispersion, age, and metallicity. Lower panels show stars selected by age. Colour shows the fraction of stars in a bin that are of a specific age.}
    \label{1399_full}
\end{figure}
\end{landscape}
\begin{landscape}
\begin{figure}
	\includegraphics[width=0.96\linewidth]{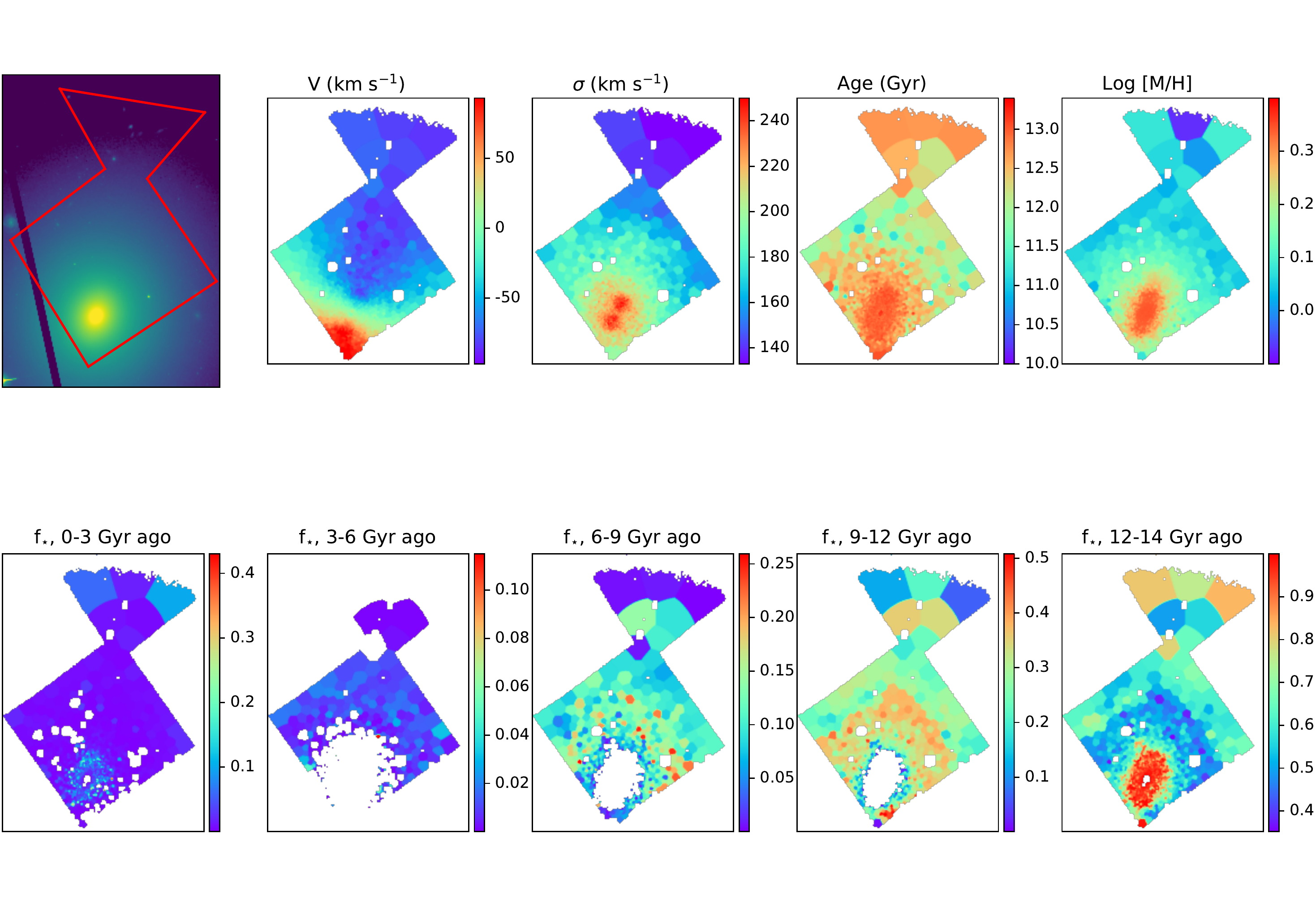}
    \caption{NGC1404. Top left panel shows a Hubble view, with the MUSE cube footprint overlayed in red. Other top row panels show stellar parameters of velocity, velocity dispersion, age, and metallicity. Lower panels show stars selected by age. Colour shows the fraction of stars in a bin that are of a specific age.}
    \label{1404_full}
\end{figure}
\end{landscape}
\begin{landscape}
\begin{figure}
	\includegraphics[width=0.96\linewidth]{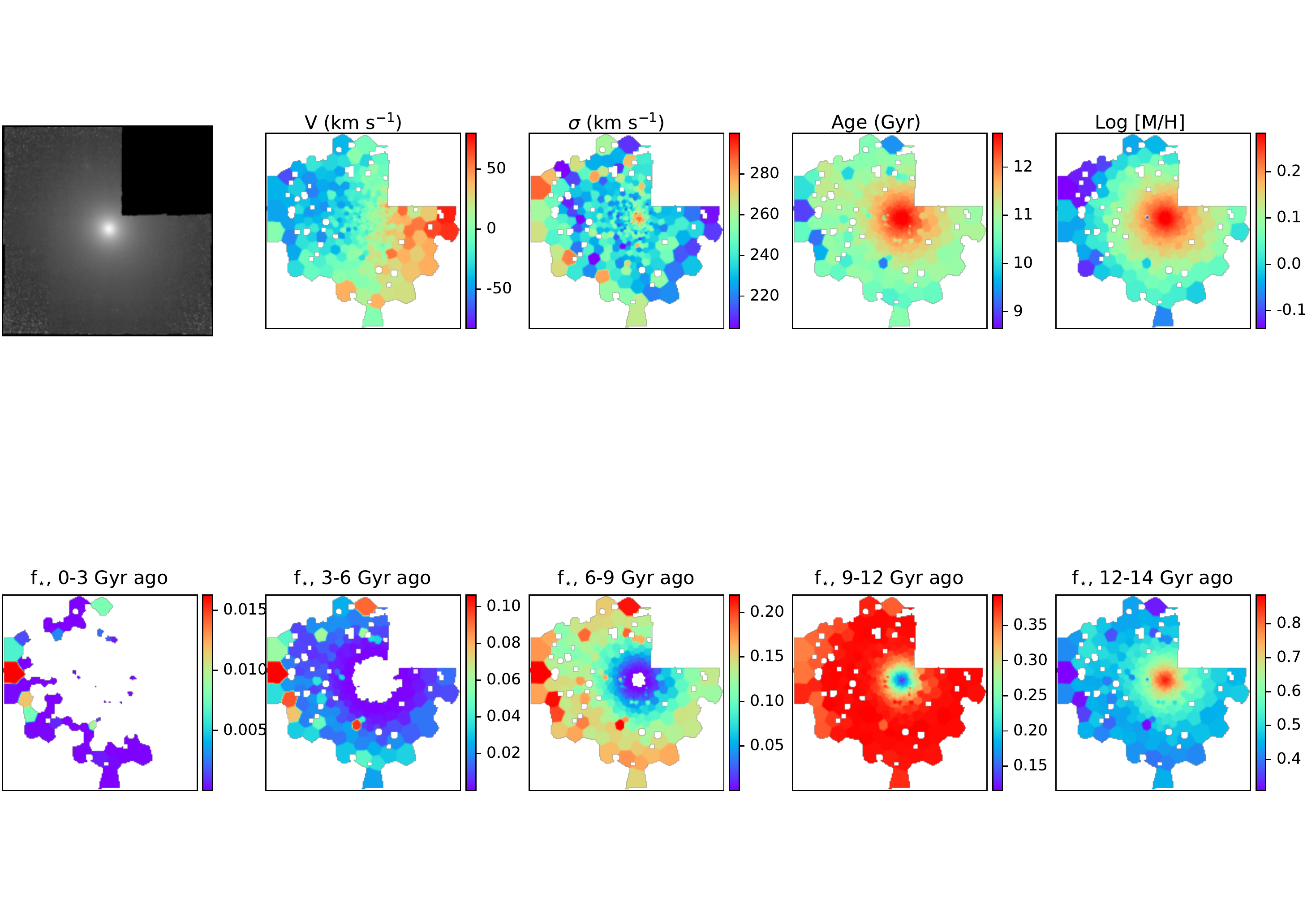}
    \caption{NGC\,1407. Top left panel shows the collapsed MUSE cube. Other top row panels show stellar parameters of velocity, velocity dispersion, age, and metallicity. Lower panels show stars selected by age. Colour shows the fraction of stars in a bin that are of a specific age.}
    \label{1407_full}
\end{figure}
\end{landscape}
\begin{landscape}
\begin{figure}
	\includegraphics[width=0.96\linewidth]{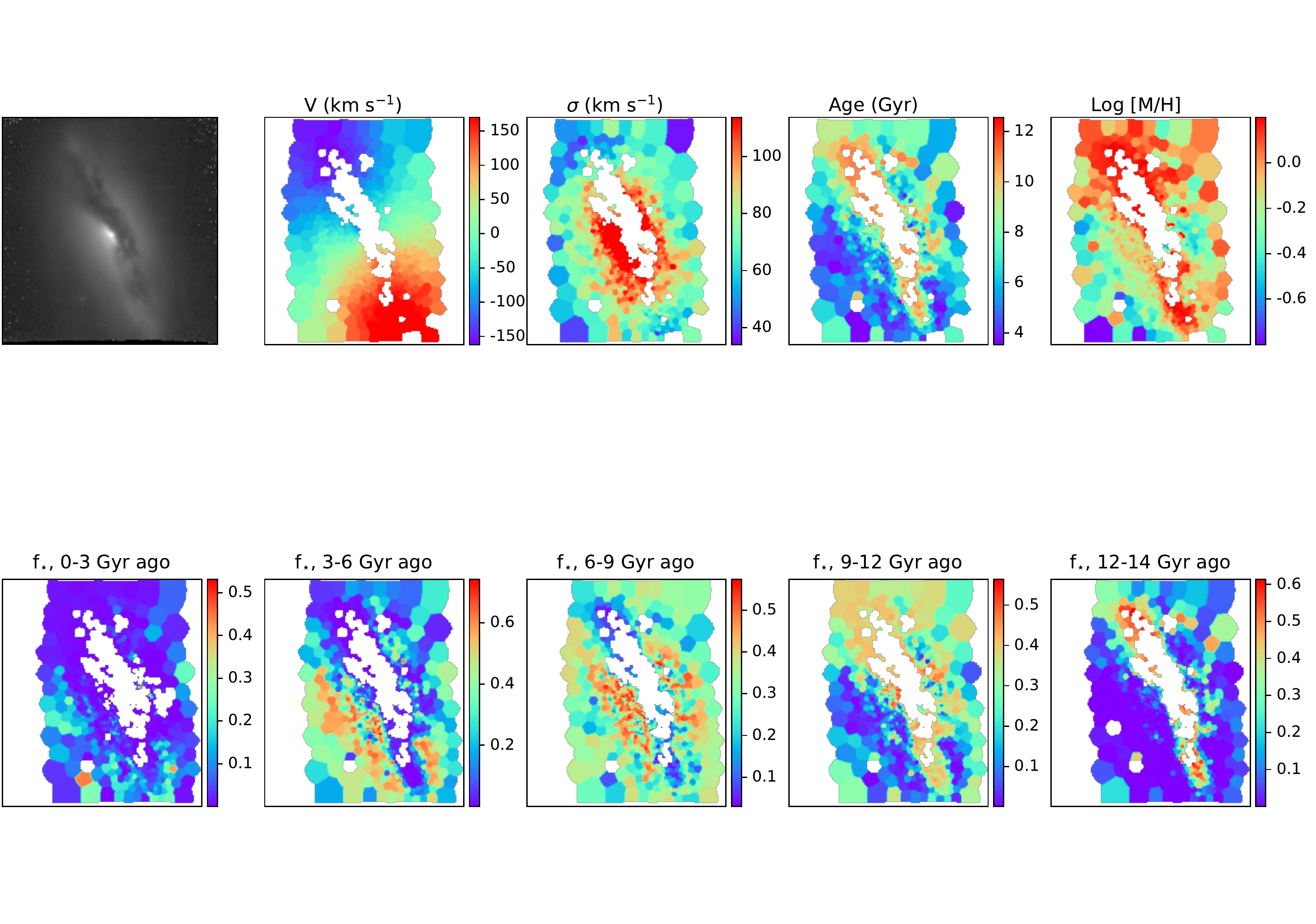}
    \caption{NGC\,2992. Top left panel shows the collapsed MUSE cube. Other top row panels show stellar parameters of velocity, velocity dispersion, age, and metallicity. Lower panels show stars selected by age. Colour shows the fraction of stars in a bin that are of a specific age.}
    \label{2992_full}
\end{figure}
\end{landscape}
\begin{landscape}
\begin{figure}
	\includegraphics[width=0.96\linewidth]{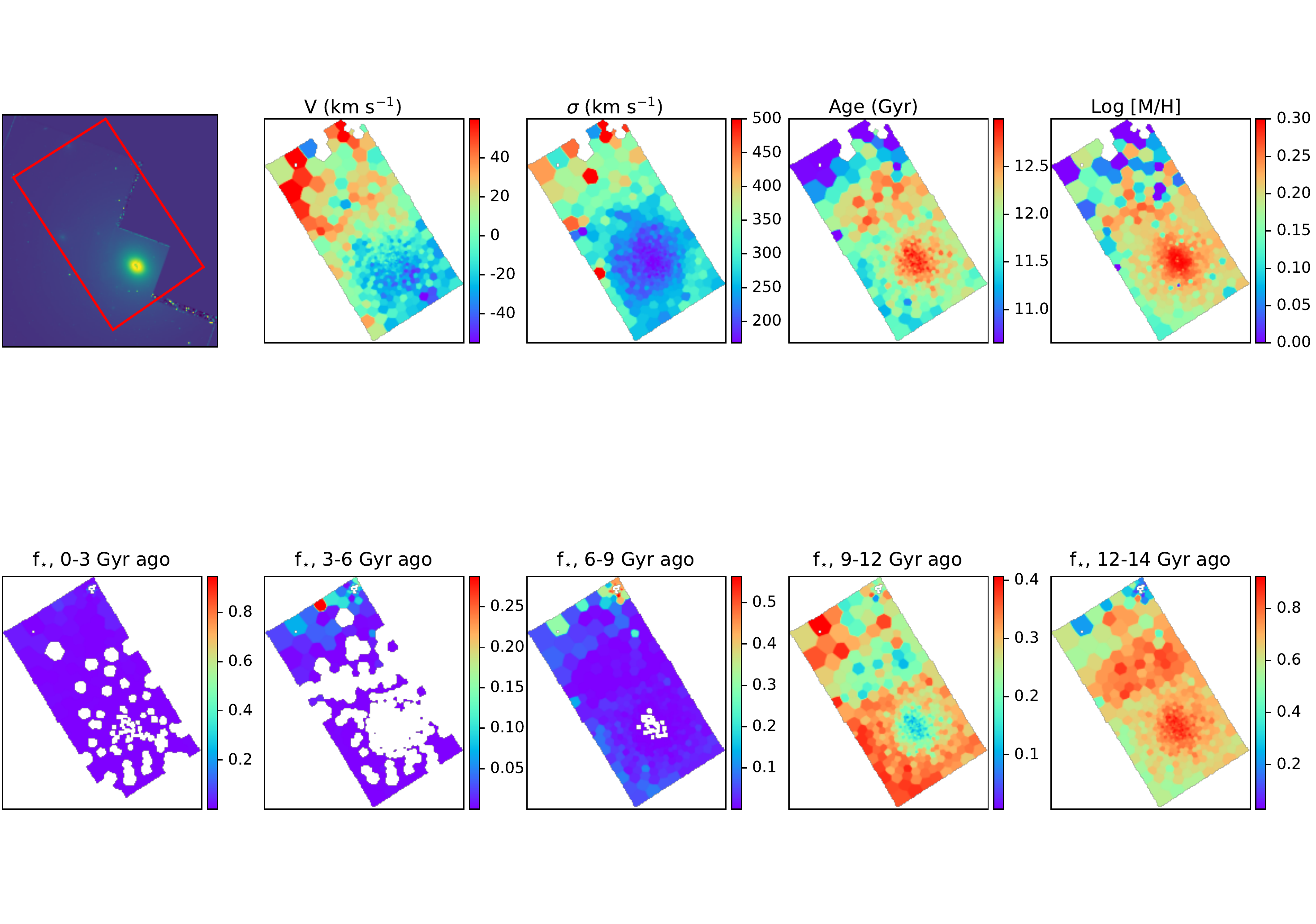}
    \caption{NGC\,3311. Top left panel shows a Hubble view, with the MUSE cube footprint overlayed in red. Other top row panels show stellar parameters of velocity, velocity dispersion, age, and metallicity. Lower panels show stars selected by age. Colour shows the fraction of stars in a bin that are of a specific age.}
    \label{3311_full}
\end{figure}
\end{landscape}
\begin{landscape}
\begin{figure}
	\includegraphics[width=0.96\linewidth]{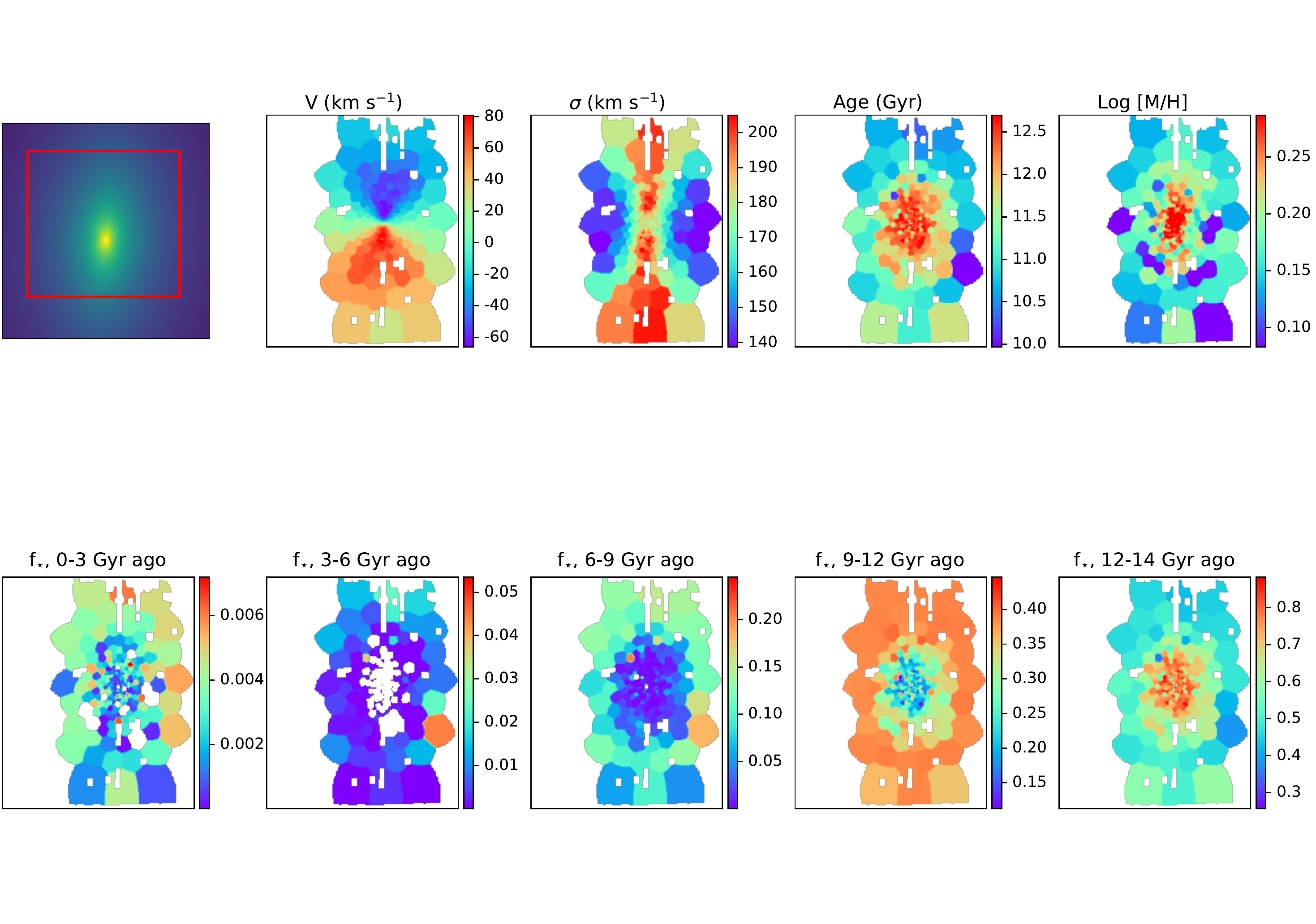}
    \caption{NGC\,4473. Top left panel shows a Hubble view, with the MUSE cube footprint overlayed in red. Other top row panels show stellar parameters of velocity, velocity dispersion, age, and metallicity. Lower panels show stars selected by age. Colour shows the fraction of stars in a bin that are of a specific age.}
    \label{4473_full}
\end{figure}
\end{landscape}
\begin{landscape}
\begin{figure}
	\includegraphics[width=0.96\linewidth]{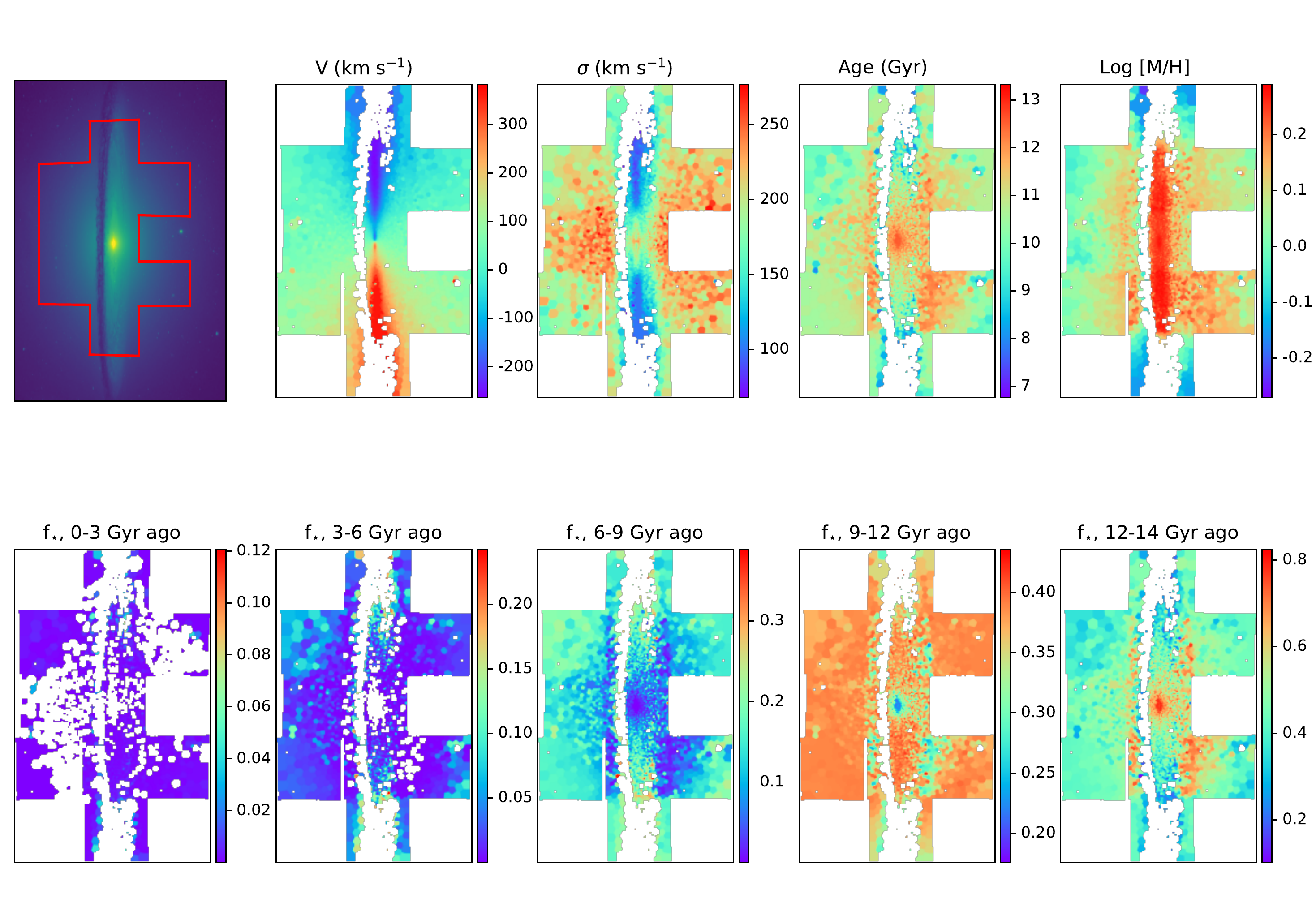}
    \caption{NGC4594. Top left panel shows a Hubble view, with the MUSE cube footprint overlayed in red. Other top row panels show stellar parameters of velocity, velocity dispersion, age, and metallicity. Lower panels show stars selected by age. Colour shows the fraction of stars in a bin that are of a specific age.}
    \label{4594_full}
\end{figure}
\end{landscape}
\begin{landscape}
\begin{figure}
	\includegraphics[width=0.96\linewidth]{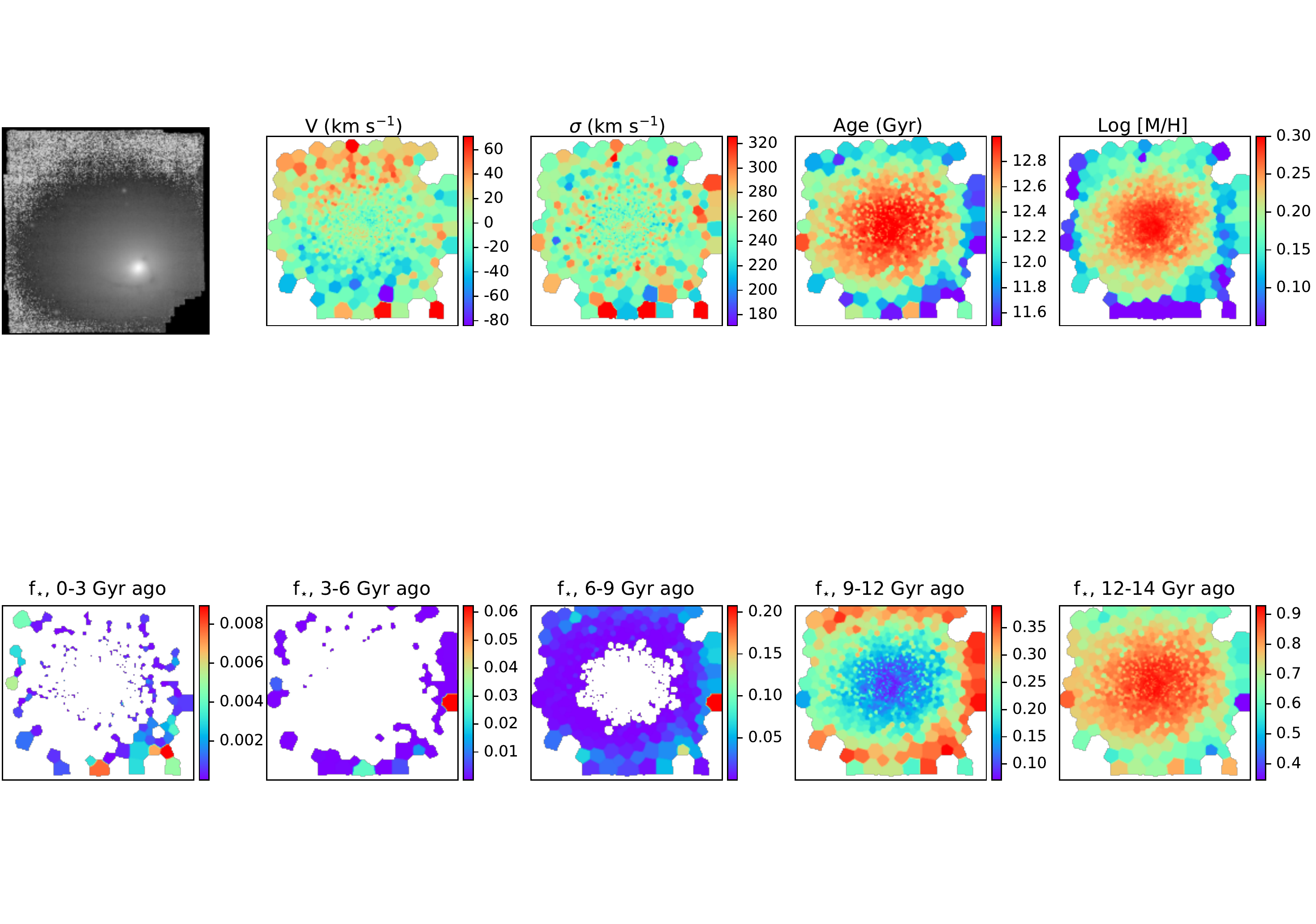}
    \caption{NGC4696. Top left panel shows the collapsed MUSE cube. Other top row panels show stellar parameters of velocity, velocity dispersion, age, and metallicity. Lower panels show stars selected by age. Colour shows the fraction of stars in a bin that are of a specific age.}
    \label{4696_full}
\end{figure}
\end{landscape}
\begin{landscape}
\begin{figure}
	\includegraphics[width=0.96\linewidth]{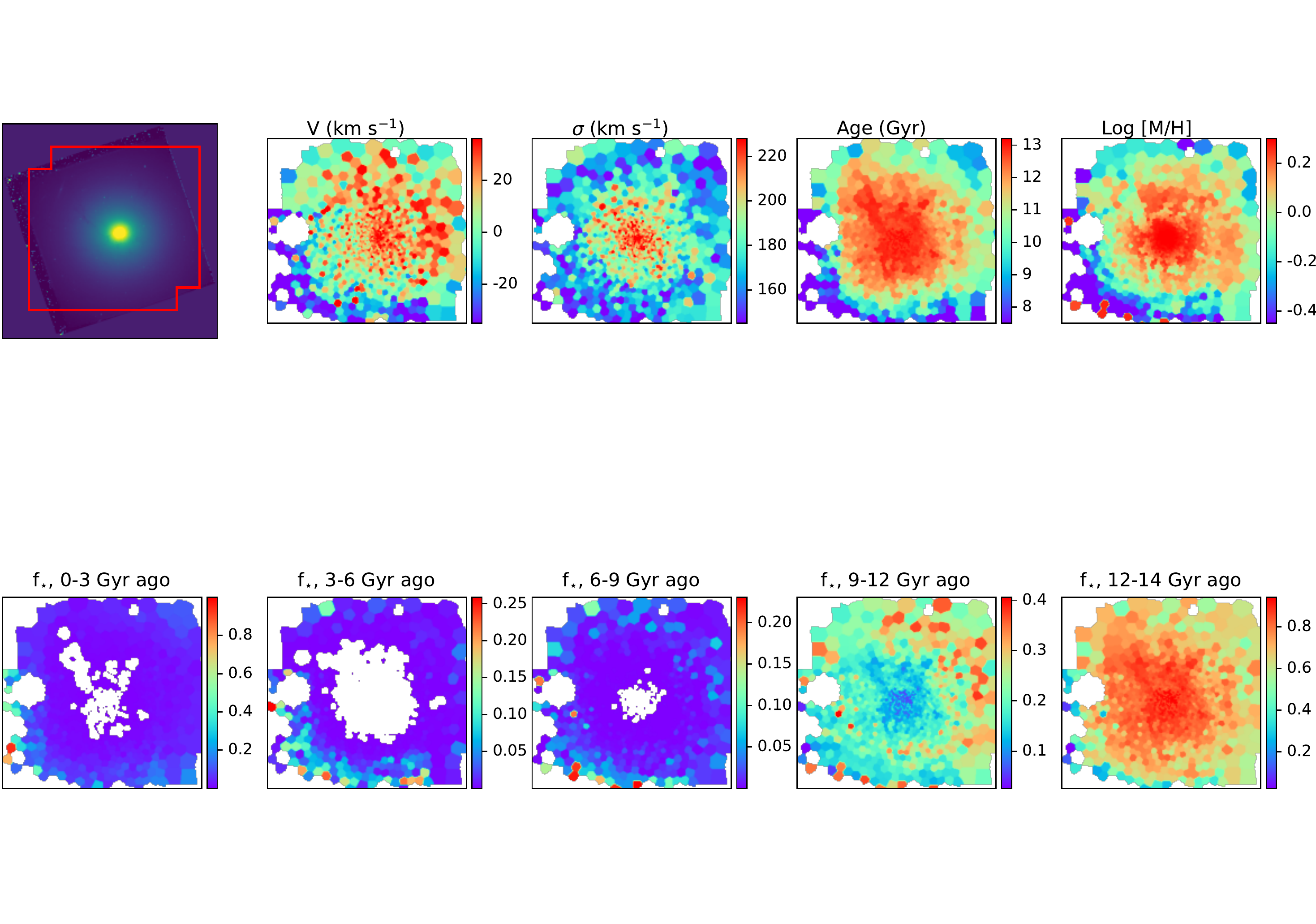}
    \caption{NGC5846. Top left panel shows a Hubble view, with the MUSE cube footprint overlayed in red. Other top row panels show stellar parameters of velocity, velocity dispersion, age, and metallicity. Lower panels show stars selected by age. Colour shows the fraction of stars in a bin that are of a specific age.}
    \label{5846_full}
\end{figure}
\end{landscape}
\let\clearpage\relax


\bsp	
\label{lastpage}
\end{document}